\documentclass[10pt,letterpaper,dvipdfmx]{article}

\usepackage[top=0.85in,bottom=1in,left=1in,right=1in,footskip=0.75in]{geometry}

\usepackage{amsmath,amssymb}

\usepackage{changepage}

\usepackage{textcomp,marvosym}

\usepackage{cite}

\usepackage{nameref,hyperref}

\usepackage[nopatch=eqnum]{microtype}
\DisableLigatures[f]{encoding = *, family = * }

\usepackage[table]{xcolor}

\usepackage{array}

\newcolumntype{+}{!{\vrule width 2pt}}

\newlength\savedwidth

\raggedright
\usepackage[aboveskip=1pt,labelfont=bf,labelsep=period,justification=raggedright,singlelinecheck=off]{caption}

\makeatletter
\renewcommand{\@biblabel}[1]{\quad#1.}
\makeatother

\usepackage{lastpage,fancyhdr,graphicx}
\usepackage{epstopdf}
\fancyheadoffset[L]{0.25in}
\fancyfootoffset[L]{0.25in}
\fancyheadoffset[R]{0.25in}
\fancyfootoffset[R]{0.25in}
\usepackage{setspace} 
\usepackage{ulem}
\usepackage{color}
\usepackage{algorithm}
\usepackage{algorithmic}
\usepackage{dcolumn}
\usepackage{bm}
\usepackage{chngcntr}

\newcommand{\grad}[0]{\mathrm{grad}}

\renewcommand{\eqref}[1]{Eq.~\ref{#1}}    

\begin{document}
\vspace*{0.2in}

\begin{flushleft}
{\Large
\textbf\newline{Network inference applicable to both synchronous and desynchronous systems from oscillatory signals} 
}
\newline
\\
Akari Matsuki\textsuperscript{1,2*},
Hiroshi Kori\textsuperscript{3},
and Ryota Kobayashi \textsuperscript{3,4\dag}
\\
\bigskip
\textbf{1} Quantitative Life Sciences, The Abdus Salam International Centre for Theoretical Physics (ICTP)
Trieste, Italy
\\
\textbf{2} Faculty of Advanced Life Science, Hokkaido University,  Sapporo, Japan
\\
\textbf{3} Graduate School of Frontier Sciences, The University of Tokyo, Chiba, Japan
\\
\textbf{4} Mathematics and Informatics Center, The University of Tokyo, Tokyo, Japan 
\\
\bigskip

* amatsuki@ictp.it \\
\dag r-koba@k.u-tokyo.ac.jp

\end{flushleft}

\date{\today}


\section*{Abstract}
Synchronization is ubiquitous in nature, which is mathematically described by coupled oscillators. Synchronization strongly depends on the interaction network, and the network plays a crucial role in controlling the dynamics. 
To understand and control synchronization dynamics in the real world, it is essential to identify the network from the observed data. 
While previous studies have developed the methods for inferring the network of asynchronous systems, it remains challenging to infer the network of well-synchronized oscillators. 
In this study, we develop a method for inferring the network of synchronized and desynchronized oscillators from time series. 
Our method expands the applicability of network inference to a wider class of oscillatory systems.
The proposed method discards a large part of data used for inference, which may seem counterintuitive. However, the effectiveness of the method is supported by the phase reduction theory, a well-established theory for weakly coupled oscillators. 
We verify the proposed method by applying it to simulated data of the limit-cycle oscillators. 
This study provides an important step towards understanding synchronization in real-world systems from a network perspective. 

\section*{Author summary}
Synchronization is an emergent phenomenon observed in populations of dynamically interacting units, which plays a crucial role across various systems, including physical, biological, chemical, engineering, and social domains. 
The network topology and the strength of coupling between elements significantly influence synchronization properties. To understand synchronization dynamics in real-world systems, it is essential to identify the interaction network from observed data. 
In this study, we propose a novel method for inferring the interaction network from oscillatory signals, which is based on the phase reduction theory for weakly coupled oscillators. Our method extends the applicability of network inference to a broader class of oscillatory systems. While the proposed method discards a substantial portion of the data, it enables accurate inference even when oscillators are highly synchronized, a situation that poses considerable challenges for existing methods. The effectiveness of the proposed method is demonstrated for a range of synthetic data, from simple phase oscillator models to biologically realistic clock cell models. This study represents an important step towards understanding synchronization mechanisms in real-world systems from a network perspective.


\section{Introduction}
\label{sec:introdution}

Synchronization is an emergent phenomenon in a population of dynamically interacting units, which plays an important role in a range of systems, including physical, biological, chemical, engineering, and social systems~\cite{pikovsky2003synchronization}. 
Synchronization phenomena are mathematically described by coupled oscillators, where an individual oscillator is modeled as a limit-cycle oscillator~\cite{winfree1967biological,kuramoto1984chemical}. 
Phase reduction theory~\cite{kuramoto1984chemical, winfree2001geometry,Nakao2016,ermentrout2019recent} is a powerful framework for analyzing the synchronization of the coupled oscillators. This theory provides a unified mathematical description of the dynamics of coupled oscillators by representing the state of the oscillator using a single variable of the phase and describing their dynamics using a reduced phase model. 
Theoretical studies based on the phase model have elucidated the key components underlying synchronization phenomena~\cite{winfree1967biological,kuramoto1984chemical,pikovsky2003synchronization,winfree2001geometry,Nakao2016,ermentrout2019recent}.

Synchronization of coupled oscillators depends strongly on the interaction network, which describes how the oscillators interact with each other. 
The structure as well as the weight of the network influence the synchronizability of the oscillators~\cite{boccaletti2006complex,arenas2008synchronization}. 
Furthermore, the network can even qualitatively change the synchronization properties. 
For the majority of networks, the Kuramoto model exhibits continuous synchronization transitions: 
the system starts to synchronize when the interaction strength reaches a threshold. 
However, the threshold is dramatically decreased in the presence of hub nodes and even vanishes in scale-free networks ~\cite{ichinomiya2004frequency, restrepo2005onset}.
It is also known that when the structural and the dynamical properties are correlated, the synchronization transition becomes discontinuous (also known as ``explosive synchronization'')~\cite{gomez2011explosive,rodrigues2016kuramoto}. 
Moreover, the network plays a central role in controlling the synchronization dynamics. For example, the control of abnormal synchronization is closely related to the treatment of brain disorders such as Parkinson's disease~\cite{popovych2014control,asllani2018minimally}. 
In addition, the network structure allows the evaluation of controllability, i.e. the ability of the system to be driven to any desired state~\cite{liu2011controllability}. Controllability has been extensively studied in various systems,
including internet, transportation, foob web, and the human brain~\cite{yuan2013exact,gu2015controllability}.

To understand and control synchronization dynamics in the real world, it is essential to identify interaction networks. 
Although advances in measurement technology have enabled us to observe the dynamics of individual oscillators, direct measurement of networks remains difficult. Therefore, it is necessary to infer the network from the observed oscillatory signals~\cite{timme2014revealing, stankovski2017coupling, turnbull2018connectivity, rosenblum2023inferring}. A variety of network inference methods have been developed for desynchronized systems~\cite{rosenblum2001detecting, paluvs2003direction, miyazaki2006determination, tokuda2007inferring, kralemann2007uncovering, cadieu2010phase,kralemann2011reconstructing, stankovski2012inference, stankovski2015coupling, ota2020interaction}. However, it remains challenging to infer the network of well-synchronized systems where the oscillators are nearly phase-locked~\cite{rosenblum2001detecting, tokuda2007inferring, tokuda2019practical}. 

To address this challenge, we propose a method for inferring the network from oscillatory signals that achieves high accuracy regardless of synchrony or asynchrony. The proposed method is based on the circle map, which describes a broad class of weakly-coupled oscillators. Our method extends the applicability of network inference to a broader class of oscillatory systems. A key feature of the method is the intentional exclusion of a substantial amount of data during the inference process. While it may seem counterintuitive to discard data in order to improve accuracy, the effectiveness of this method is supported by phase reduction theory.
To test the validity of the proposed method, we apply it to simulated data from two realistic limit-cycle oscillator models: the Brusselator for chemical oscillators and the model of circadian oscillators in the suprachiasmatic nucleus (SCN). 

\noindent

\section{Results}\label{sec2}
\subsection{Phase description of coupled oscillators}
\label{sec:phase-desc}
Phase reduction theory is a mathematical framework that provides a simplified description of weakly coupled nonlinear limit-cycle oscillators in general. 
We consider a system of $N$ weakly coupled limit-cycle oscillators. 
The dynamics of oscillator $i$ ($i=1,\ldots,N$) can be written as 
\begin{eqnarray}
    \frac{d {\bm X}_i}{dt} =
{\bm F}({\bm X}_i)+ \epsilon \bm f_i({\bm X}_i)
+ \epsilon \sum_{j=1}^N {\bm Q}_{ij} \left({\bm X}_i, {\bm X}_j \right) + {\bm \eta_i(t)}, 	 \label{eq:dyneq-noisy-coupled-osc}
\end{eqnarray}
where 
$\bm F({\bm X}_i)$ represents the unperturbed dynamics of a typical oscillator, 
$\epsilon \bm f_i({\bm X}_i)$ represents the difference of the intrinsic dynamics of oscillator $i$ from the typical oscillator, and $\epsilon {\bm Q}_{ij}({\bm X}_i, {\bm X}_j)$ represents the interaction from oscillator $j$ to oscillator $i$. 
Note that $\epsilon$ is a dimensionless parameter that characterizes the degree of heterogeneity and the coupling strength (i.e., the interaction strength).  
The noise term ${\bm \eta_i(t)}$ is assumed to be the Gaussian white noise obeying ${\rm E}[ {\bm \eta_i(t)}]= {\bm 0}$ and ${\rm Cov}[ {\bm \eta_i(t)} {\bm \eta_i(s)}]= v_i^2 I_d \delta(t-s)$, 
where $I_d$ is the identity matrix of size $d$, $v^2_i$ is the noise variance, and $\delta(t)$ is the Dirac's delta function.  

The dynamics of the system (\eqref{eq:dyneq-noisy-coupled-osc}) can be accurately described in terms of the phase $\phi_i$ of each oscillator if the perturbations to the limit-cycle oscillator 
are sufficiently small, i.e., $\epsilon, v_i \ll 1$. 
%
Specifically, the following phase equation is derived from \eqref{eq:dyneq-noisy-coupled-osc} by neglecting 
$O(\epsilon^2)$ and $O(v_i^2)$ terms (see Sec.~\ref{method:derive-asympt-phase-eq}):
\begin{eqnarray}
    \frac{d\phi_i}{dt} = \omega + \epsilon \nu_i(\phi_i)  + \epsilon \sum_{j=1}^N q_{ij}(\phi_i, \phi_j) + {\bm Z}(\phi_i) \cdot {\bm \eta}_i(t),
    \label{eq:phase-eq-nonaveraged}
\end{eqnarray}  
where ${\bm Z}(\phi)$ is the phase sensitivity function, $q_{ij}(\phi_i, \phi_j)$ is the coupling function. 
Thus, the $dN$-dimensional dynamical system, given by \eqref{eq:dyneq-noisy-coupled-osc}, is reduced to the $N$-dimensional system. 
Moreover, by applying the averaging approximation~\cite{kuramoto1984chemical, Nakao2016,ermentrout2019recent}, we can derive a simpler equation
\begin{eqnarray}
    \frac{d \phi_i}{dt} =	\omega_i  + \sum_{j=1}^N c_{ij} \gamma_{ij}( \phi_j- \phi_i ) + \sigma_i \xi_i(t),     
    \label{eq:noisy-kuramoto-coupling-averaged}
\end{eqnarray}
where $\omega_i$ is the natural frequency of oscillator $i$, $c_{ij}$ is the coupling strength from oscillator $j$ to $i$, $\gamma_{ij}(\phi)$ is the averaged coupling function, $\sigma_i^2$ is the  noise variance, and $\xi_i(t)$ is the Gaussian white noise with the mean $0$ and the variance $1$. 
Note that the averaged model (\eqref{eq:noisy-kuramoto-coupling-averaged}) is a less accurate description of the original system (\eqref{eq:dyneq-noisy-coupled-osc}) than the non-averaged model (\eqref{eq:phase-eq-nonaveraged}): there is a discrepancy of $O(\epsilon)$ between \eqref{eq:phase-eq-nonaveraged} and \eqref{eq:noisy-kuramoto-coupling-averaged} (see Sec.~\ref{method:derive-asympt-phase-eq} for details).

\subsection{Proposed method for inferring the coupling network of oscillators}
\label{sec:proposed}

\begin{figure}[t]
    \centering
    \includegraphics[width=\hsize]{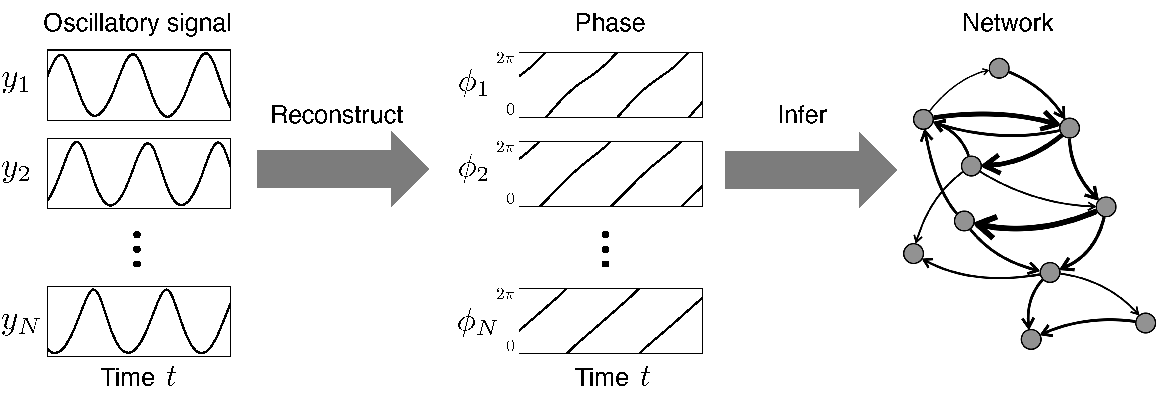}
    \caption{
    Procedure for network inference of an oscillator system. 
    (i) Reconstructing the phase signals $\phi_i(t)$ from the observed signals $y_i(t)$. 
    (ii) Inferring the interaction network from the phase signals. 
    The direction and the width of the arrow represent the direction and the weight of the network, respectively.}     
    \label{fig:inference-protocol}
\end{figure}

Our goal is to infer the interaction network or coupling network, 
i.e. the coupling strengths among oscillators, from the oscillatory signals. 
As illustrated in Fig ~\ref{fig:inference-protocol}, we first reconstruct the phase $\phi_i(t)$ of each oscillator from the observation $y_i(t)$ and fit the reconstructed phases to a model. 
Thanks to the phase reduction theory (see Sec.~\ref{sec:phase-desc}), our approach via phase reconstruction is generally applicable to arbitrary limit-cycle oscillators with weak heterogeneity, coupling, and noise. 
Moreover, it is not necessary to infer the detailed properties of the oscillators, since we are only interested in identifying the interaction network. 
Therefore, it makes sense to consider a phase oscillator model as a model for inferring the network. 

Our proposed method is based on the circle map~\cite{pikovsky2003synchronization}, which describes the evolution of the phase over a period of an oscillator.  
Consider the change in phase $\phi_i(t)$ of oscillator $i$ from time $t$ to $t+T$, where $T= 2\pi/ \omega$ represents the typical period of an oscillator.
We can derive the circle map from the phase equation (\eqref{eq:phase-eq-nonaveraged}) under the assumption of weak coupling $O(\epsilon)$  (see \nameref{appendix:derive-cm} for details): 
\begin{eqnarray}
	\phi_i(t+T)- \phi_i(t)= T \omega_i + T \sum_{j=1}^N c_{ij} \gamma_{ij}\left( \phi_j(t) -\phi_i(t) \right) +  \sqrt{T}\sigma_i \xi_{i, t}  + O(\epsilon^2),	 \label{eq:circle-map-cont}
\end{eqnarray}
where $N$ is the number of oscillators, $c_{ij}$ is the coupling strength from oscillator $j$ to $i$, 
$\gamma_{ij}(\phi_j(t) -\phi_i(t))$ is the averaged coupling function, 
$\sigma^2_i$ is the averaged noise variance,   
and $\xi_{i, t}$ are the independent Gaussian random variables with mean $0$ and variance $1$. 
The error between the original model (\eqref{eq:dyneq-noisy-coupled-osc}) and the circle map (\eqref{eq:circle-map-cont}) is $O(\epsilon^2)$. 
This argument implies that the circle map is a superior approximation to the averaged equation (\eqref{eq:noisy-kuramoto-coupling-averaged}), exhibiting a discrepancy of $O(\epsilon)$ from the original model. 
The circle map captures the dynamics while neglecting faster dynamics relative to the oscillation period. Consequently, the proposed method, which is based on the circle map, does not require data with a small sampling interval and can discard a large portion of the data.

It is challenging to infer the interaction network from the observed data when the oscillators are partially synchronized or in a nearly phase-locked state~\cite{rosenblum2001detecting, tokuda2007inferring, tokuda2019practical}. 
This difficulty arises because the amount of information available to infer the coupling is limited when the oscillators are synchronized.
To reduce the difficulty, we make two assumptions for the network inference. 
First, we assume that the coupling function is the same for all the pairs, $\gamma_{ij}(\phi_j- \phi_i)= \gamma(\phi_j- \phi_i)$. 
Second, the averaged coupling function is assumed to be a sinusoidal function containing only the first harmonic term~\cite{sakaguchi1986soluble}. 
This assumption is reasonable when the oscillation is close to the Hopf bifurcation~\cite{kuramoto1984chemical,Kori2014} or the oscillators are nearly phase-locked state. 
Therefore, we assume
\begin{eqnarray}
    \gamma_{ij}( \phi_j - \phi_i )= \sin(\phi_j -\phi_i + \alpha),    \label{eq:gamma}	
\end{eqnarray}
where $\alpha$ is a parameter that controls the phase-locked state. 
The circle map used for the inference is obtained by substituting the sinusoidal coupling function (\eqref{eq:gamma}) 
\begin{eqnarray}
   	\Delta \Phi_{i, m} = T \omega_i +  T \sum_{j=1}^N c_{ij}\sin\left(  \Phi_{j, m} - \Phi_{i, m} + \alpha \right)  + \sqrt{T} \sigma_i \xi_{i, m},    \label{eq:circle-map}
\end{eqnarray}
where $m$ is an integer greater than or euqal to $0$ ($m= 0, 1, 2, ..., M-1$),
$\Phi_{i, m} = \phi_i( mT )$ is the phase of oscillator $i$, $\Delta \Phi_{i, m} =  \Phi_{i, m+1} -  \Phi_{i, m}$ is the phase change over the typical period $T$, and $\xi_{i, m}$ an independent Gaussian random variable with mean 0 and variance 1. 
Here the typical period $T$ is estimated from the reconstructed phase $\{ \phi_{i,k} \}$ in advance (see procedure (ii) below).   
Finally, we obtain the network $\{c_{ij}\}$ by fitting the phase $\{ \Phi_{i,k} \}$ to \eqref{eq:circle-map} using the maximum likelihood method. 

Here, we describe the procedure of the proposed method. 
Assume that an oscillatory signal is observed at $K$ time steps, $y_{i,k} = y_i(kh)$ ($k=0,1,..., K-1$), where $h$ is the original sampling interval.
The proposed method consists of three steps.
\begin{itemize}
   \item[(i)]    Reconstruct the phase $\{  \phi_{i, k} \}$ of each oscillator from the observed data $\{ y_{i, k} \}$ $(i= 1, 2, \cdots, N; k= 0, 1, \cdots, K-1)$ by using the Hilbert transform~\cite{chavez2006,king2009hilbert-vol1,matsuki2023extended}. 
   \item[(ii)]   Estimate the typical period $T$. We first calculate the average period $ \langle  \tau \rangle = \frac{1}{N} \sum_{i=1}^N \tau_i$, where $ \tau_i= \left( \frac{ \phi_{i, K-1} - \phi_{i, 0} }{ 2\pi h (K-1)} \right)^{-1}$ is the period of oscillator $i$. 
   We then estimate the typical period $T$ by $\hat{T}= Lh$, where $L$ is an integer nearest to $\frac{\langle  \tau \rangle}{h}$. 	 
   \item[(iii)]  
    Determine the parameters in \eqref{eq:circle-map} including the coupling strength $\{ c_{ij} \} \ (i, j=1,2,...,N)$ from the reconstructed phase $\{ \phi_{i,k} \}$ using the maximum likelihood method (see Sec \ref{method:mle} for details). Note that we do not infer the self-coupling $c_{ii}$, since it is assumed to be zero, $c_{ii} = 0\ (i=1,2,...,N)$.    
\end{itemize}

\subsection{Simple examples: A system of two phase oscillators }
In this subsection, we focus on a system of two phase oscillators and examine the performance of the proposed method. Here, the proposed method is apppied to two phase oscillators, i.e., Kuramoto model and Winfree model. 

\subsubsection{Kuramoto model }
\begin{figure}
    \centering
    \includegraphics[width=.8\hsize]{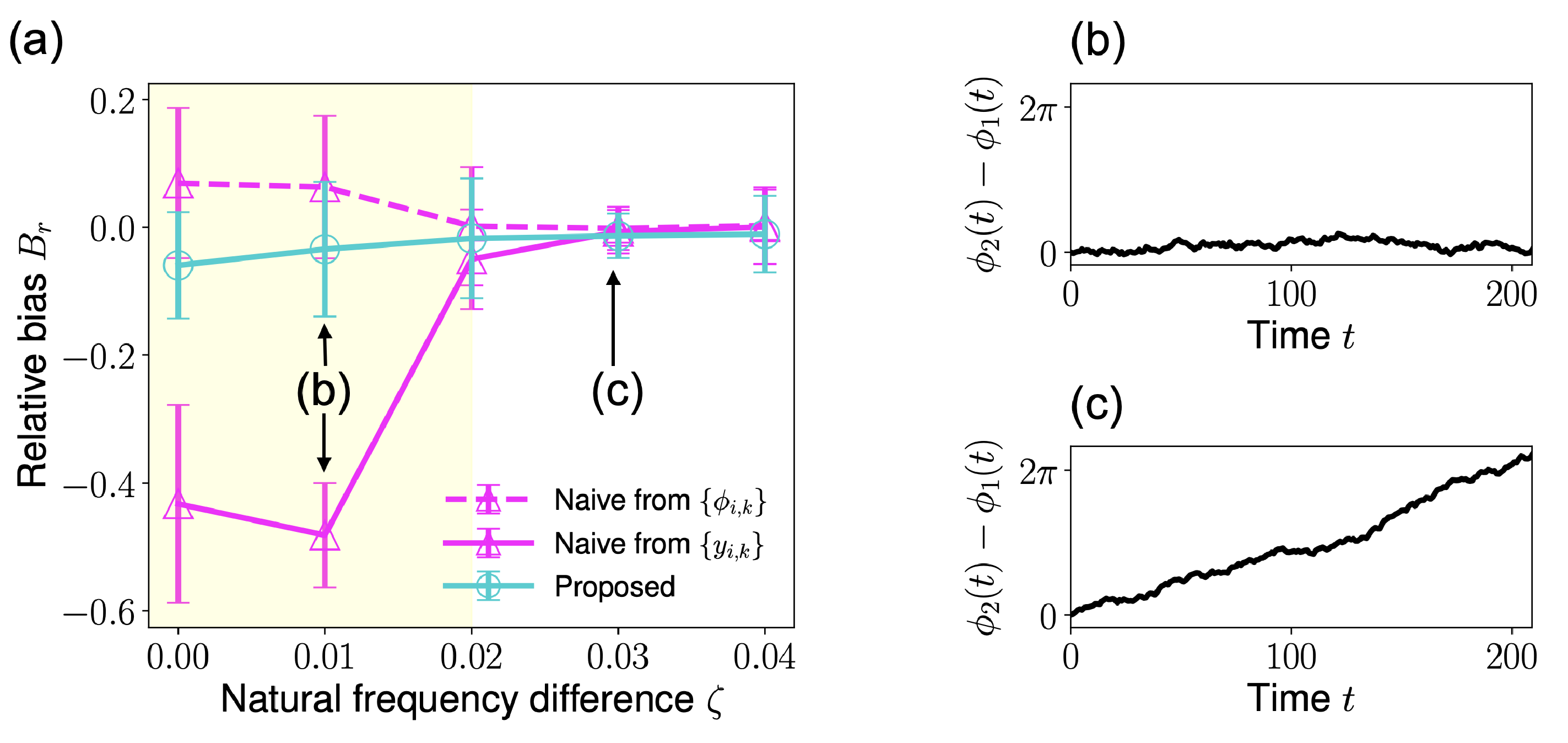}
    \caption{ 
    Inference performance for a system of two phase oscillators: Kuramoto model. 
    (a): Mean and standard deviation of the relative bias of the inferred coupling strength. 
    The cyan and magenta solid lines represent the performance of the naive method (see \nameref{appendix:naive-method}) and the proposed method, respectively. The dashed magenta line represents the performance of the naive method when the phase is directly observed. The yellow region indicates the region where the oscillators are synchronized in the absence of noise. 
    (b) and (c): Time series of phase differences between two oscillators in synchronous (b) and asynchronous (c) states. 
    Model parameters are set as $\omega_1=1.0$, $c= 0.01$, $\sigma_1 = \sigma_2=0.01$, and the natural frequency difference $\omega_2 - \omega_1 = 0.0$ (b) and $0.04$ (c).   
    }
    \label{fig:method1}
\end{figure}

We evaluate the inference performance of the proposed method in the case of a simple phase oscillator model. Specifically, we focus on a system of two oscillators that are coupled bidirectionally: 
\begin{eqnarray}
     \frac{d\phi_1}{dt} &=& 	\omega_1 + c  \sin (\phi_2- \phi_1) +  \sigma_1  \xi_1(t),	\label{eq:noisy-KS1}		\\
	\frac{d\phi_2}{dt} &=& 	\omega_2 + c  \sin (\phi_1- \phi_2) +  \sigma_2  \xi_2(t),	\label{eq:noisy-KS2}
\end{eqnarray}  
where $\omega_{1, 2}$ is the natural frequency, $c$ is the coupling strength, and $\sigma_{1, 2}$ represents the noise strength. 
This model (Eqs. \ref{eq:noisy-KS1}, \ref{eq:noisy-KS2}) is a stochastic Kuramoto model \cite{pikovsky2003synchronization}, which is an averaged model (\eqref{eq:noisy-kuramoto-coupling-averaged}) with the sinusoidal coupling function, $\gamma_{ij}( \phi_j - \phi_i )= \sin(\phi_j -\phi_i)$. 
The synthetic data $\{\phi_{i,k}\}$ ($i= 1,2$; $k= 0, 1, \cdots, K-1$) (total duration: $Kh= 20,000$ in Fig ~\ref{fig:method1}) are generated by applying the Euler-Maruyama method, i.e., the Euler method for stochastic differential equations~\cite{kloeden1992stochastic}, to Eqs. \ref{eq:noisy-KS1} and \ref{eq:noisy-KS2} with a time step of $0.01$. The synthetic data is generated from the phase time series: $y_{i,k} = \cos\phi_{i,k}$. 
The inference performance is evaluated based on the relative bias defined as $B_r= (\hat{c}_{12} - c_{12})/c_{12}$, where $c_{12}$ and $\hat{c}_{12}$ are the true coupling strength from oscillator $2$ to $1$ and its inferred value, respectively.

We examine the effect of synchronization on inference performance. In the Kuramoto model, the synchronization state can be controlled by adjusting the natural frequency difference $\zeta := \omega_2 - \omega_1$. Let us consider the Kuramoto model (Eqs. \ref{eq:noisy-KS1}, \ref{eq:noisy-KS2}) with positive coupling ($c > 0$) and no external noise ($\sigma_1 = \sigma_2 = 0$). 
The synchronization state is determined by the natural frequency difference $\zeta$. Specifically, the two oscillators are synchronized or phase-locked if the difference in natural frequencies is less than a critical value: $|\zeta| < 2c$.

Fig ~\ref{fig:method1}(a) compares the performance of the proposed method using the synthetic signal $\{ y_{i,k} \}$ $(i= 1, 2; k= 0, 1, ..., K-1)$ with that of the naive method 
that infers the interaction network by fitting parameters of the averaged model (\eqref{eq:noisy-kuramoto-coupling-averaged}) (see \nameref{appendix:naive-method} for details).
The proposed method demonstrates excellent performance across a broad range of natural frequency differences $\zeta$, indicating its capability to accurately infer coupling strength in both synchronous and asynchronous regimes. 
In contrast, the naive method performs well in the asynchronous regime ($\zeta = 0.02, 0.03$, and $0.04$), but it considerably underestimates the coupling strength in the synchronous regime ($\zeta = 0.00, 0.01$). 
The oscillators exhibited synchronous (asynchronous) activity when the natural frequency differences were less (greater) than the critical value (Fig ~\ref{fig:method1}b and \ref{fig:method1}c). 
When the phase signal $\{ \phi_{i,k} \}$ is observed directly,
the naive method performs well even in the synchronous regime (Fig ~\ref{fig:method1}a, dashed magenta). 
This suggests that the phase reconstruction has a significant negative impact on coupling inference based on the naive method when oscillators are synchronized. 
Next, we briefly discuss why phase reconstruction in the naive method negatively affects coupling inference. When the oscillators synchronize, the phase change fluctuates rapidly because the coupling term $\sin(\phi_{2,k} - \phi_{1,k})$ is nearly constant (Fig ~\ref{fig:method1}b) and the noise term $\xi_{i,k}$ dominates the phase change $\delta \phi_{i,k} = \phi_{i,k+1} - \phi_{i,k}$. Due to the properties of the Hilbert transform \cite{matsuki2023extended}, the rapid fluctuation in the phase change is smoothed, degrading the coupling inference.

Finally, the inference performance of the proposed method was compared with that of Convergent Cross Mapping (CCM)~\cite{sugihara2012detecting}, a method for inferring a causal relationship between time series based on nonlinear state space reconstruction. 
Here, the observed signals $\{ y_{i,k} \}$ $(i= 1, 2; k= 0, 1, ..., K-1)$ (total duration: $Kh= 20,000$ in Tables~\ref{tab:comp-ccm-1} and \ref{tab:comp-ccm-2}) are embedded in two dimensional space. The sampling sampling interval of the CCM method is set to $h=0.63$ to avoid the enormous computational cost. 
Note that the CCM method is not able to infer the coupling strength itself. Instead, it returns a value, $\rho_{ij}$, in the range of $[-1, 1]$. A value of $\rho_{ij}$ close to 1 indicates a causal relationship from oscillator $j$ to oscillator $i$.
First, we focus on the detectability of the coupling and examine the effect of synchronization on the coupling inference (Table~\ref{tab:comp-ccm-1}). 
While CCM is able to detect the coupling from synchronized oscillators ($\zeta < 0.02$), it is unable to detect the coupling from asynchronous oscillators ($\zeta > 0.02$). 
In contrast, the proposed method demonstrated an ability to accurately infer the coupling strength as shown in Fig ~\ref{fig:method1}(a). 
Second, we examine whether the method can infer the asymmetry of the coupling. Consider a system of two phase oscillators with asymmetric coupling: 
\begin{align}
    \frac{d\phi_1}{dt} &= \omega_1 + \ \ c \sin(\phi_2 - \phi_1) + \sigma_1 \xi_1(t), \\
    \frac{d\phi_2}{dt} &= \omega_2 + \gamma c \sin(\phi_1 - \phi_2) + \sigma_2 \xi_2(t),
\end{align}
where $\gamma$ represents the asymmetry of the coupling. 
The asymmetry $\gamma$ is inferred by using the proposed method and the CCM method. While the asymmetry is inferred by calculating the ratio of the coupling strength, $\hat{c}_{21} / \hat{c}_{12}$, in the proposed method, it is inferred by calculating the ratio of the causal strength, $\hat{\rho}_{21} / \hat{\rho}_{12}$, in the CCM method.    
The proposed method demonstrated an effective capability to accurately infer the asymmetry of the coupling. In contrast, CCM was unable to infer the asymmetry from time series (Table~\ref{tab:comp-ccm-2}). 
One reason for the suboptimal performance of CCM in terms of inference is that it defines causality based on predictability. This type of causality differs from the coupling strength $|Q_{ij}|$ defined in equation (\eqref{eq:dyneq-noisy-coupled-osc}), which is the target of inference in this study. The CCM method, which does not make any assumptions on specific dynamics, can be applied to chaotic systems~\cite{sugihara2012detecting}. 
However, the result (Tables~\ref{tab:comp-ccm-1} and~\ref{tab:comp-ccm-2}) suggests that the CCM method is not suitable for accurately inferring the coupling network between oscillators. 

\begin{table}[]
    \centering
    \begin{tabular}{ c | c  c  c  c  c}
        \hline
         $\zeta$ & \ 0.00 \  & \ 0.01 \  & \  0.02 \  &  \ 0.03 \ &  \ 0.04 \  \\
        \hline  \hline
         $\rho_{12}$  &  {\bf 0.92}  &  {\bf 0.89}  &  0.48  &  0.19  &  0.11  \\
         $\rho_{21}$  &  {\bf 0.92}  &  {\bf 0.89}  &  0.48  &  0.20  & 0.12           
    \end{tabular}
    \caption{
    Inferred coupling strength by CCM method. 
    As in Fig ~\ref{fig:method1}, the coupling strength was inferred from two Kuramoto oscillators with different natural frequency differences $\zeta$. The inferred results from synchronous oscillators ($\zeta < 0.02$) are shown in bold. 
    The oscillators were weakly coupled: $c_{12} = c_{21} =0.01$.
    }
    \label{tab:comp-ccm-1}
\end{table}

\begin{table}[]
    \centering
    \begin{tabular}{l  c c c}
        \hline
        Coupling ratio $\gamma$ & 0 & 0.5 & 1.0 \\
        \hline
         Proposed & {\bf 0.061} & {\bf 0.54} & {\bf 1.0} \\
         CCM & 1.0 & 1.0 & {\bf 1.0} 
    \end{tabular}
    \caption{
    Inferring the asymmetry of the coupling. 
    The inferred coupling ratio $\gamma$ obtained by the proposed method was compared with that obtained by the CCM method. The more accurate results are shown in bold. 
    }
    \label{tab:comp-ccm-2}
\end{table}
\clearpage

\subsubsection{Winfree model }
We evaluate the inference performance of the proposed method in the case of the non-averaged phase model (Eq.~\ref{eq:phase-eq-nonaveraged}).  
As the true model, we consider a system of two phase oscillators coupled bidirectionally: 
\begin{eqnarray}	
	\frac{d\phi_1}{dt} &=& \omega_1 -  2c \sin \phi_1 (1+ \cos\phi_2) +  \sigma_1 \sin \phi_1  \xi_1(t),  \label{eq:noisy-winfree1}       \\
	\frac{d\phi_2}{dt} &=& \omega_2 -  2c \sin \phi_2 (1+ \cos\phi_1) +  \sigma_2 \sin \phi_2  \xi_2(t),  \label{eq:noisy-winfree2} 
\end{eqnarray}  
where, $\omega_{1, 2}$ is the natural frequency, $c$ is the coupling strength, and $\sigma_{1, 2}$ represents noise strength.  
This model is a stochastic Winfree model~\cite{winfree1967biological, ariaratnam2001phase}, which is a special case of the non-averaged model (Eq.~\ref{eq:phase-eq-nonaveraged}):  $\epsilon  \nu_i(\phi_i)=  \omega_i-  \omega$,  $\epsilon q_{ij}(\phi_i,  \phi_j)= -  2c  \sin  \phi_i (1+ \cos \phi_j)$, and $Z(\phi_i)=  \sin  \phi_i$.
Phase time series $\{\phi_{i,k}\}$ $(i=1, 2; k= 0, 1, ..., K-1)$ (total duration: $Kh= 20,000$) are generated by simulating the true model (Eqs.~\ref{eq:noisy-winfree1} and \ref{eq:noisy-winfree2}) using the Euler-Maruyama method with a time step of $0.01$. 
Two methods (the proposed method and the naive method: see \nameref{appendix:naive-method} for details) are applied to the phase time series to infer the coupling strengths ($c_{12}$ and $c_{21}$) between oscillators. 
Note that the true phase signal is assumed to be observable to eliminate the effect of the phase reconstruction. 
Similar to the Kuramoto model, the performance is evaluated based on the relative bias $B_r$.

\begin{figure}
    \centering
    \includegraphics[width=.8\hsize]{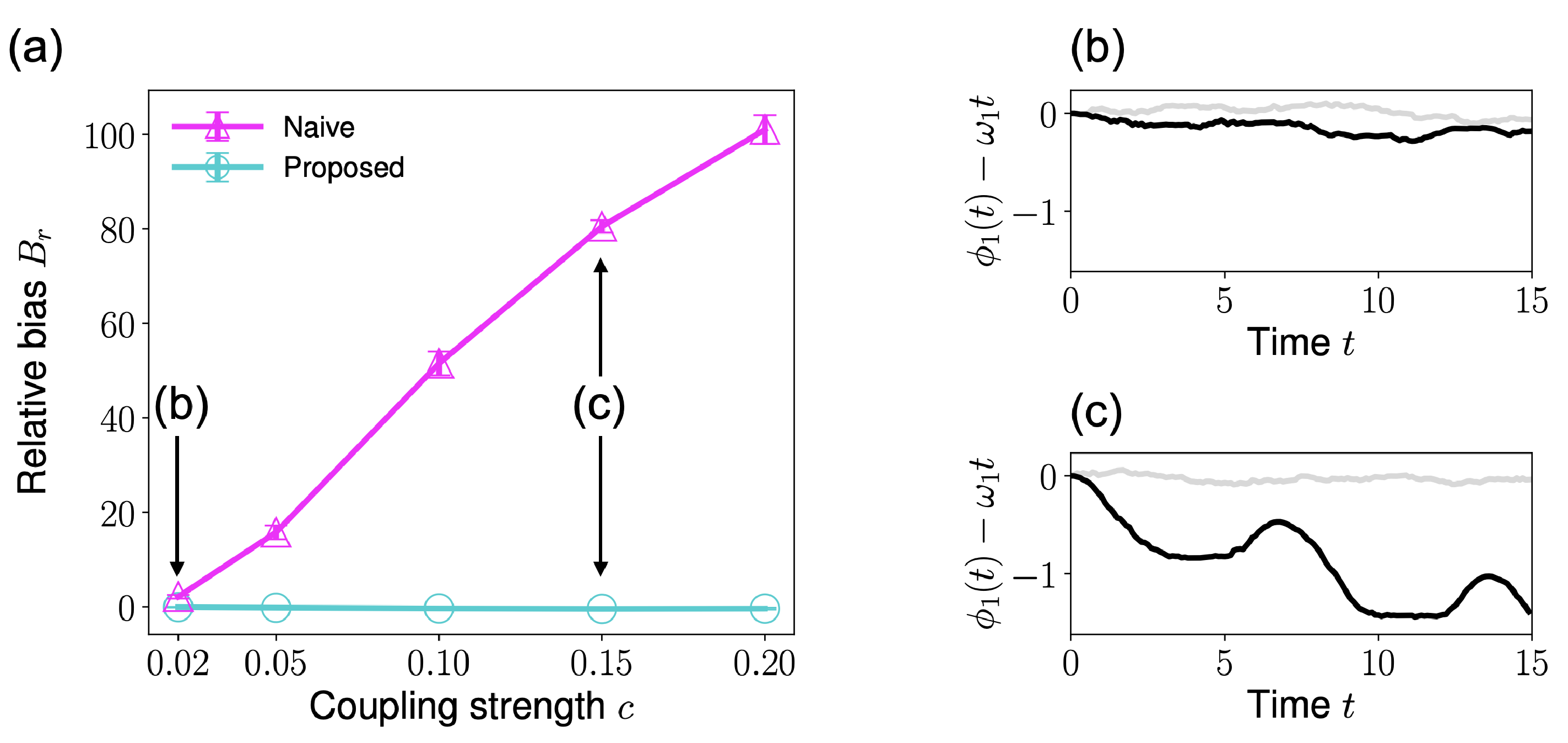}
    \caption{ 
    Inference performance for a system of two phase oscillators: Winfree model. 
    (a): Mean and standard deviation of the relative bias of the inferred coupling strength. 
    The cyan and magenta solid lines represent the performance of the proposed method and the naive methods, respectively.   
    (b) and (c): Phase time series obtained from the weak (b) and moderate (c) coupling strengths. Note that the natural frequency component $\omega_i t$ is substracted from the phase $\phi_i(t)$. Black and gray lines represent the phase of the Winfree model (Eqs.~\ref{eq:noisy-winfree1} and \ref{eq:noisy-winfree2}) and its avegared model (Eqs.~\ref{eq:noisy-KS1} and \ref{eq:noisy-KS2}), respectively. 
    Model parameters are set as $\omega_1= \omega_2= 1.0$ $\sigma_1 = \sigma_2=0.05$, and the coupling strength $c= 0.02$ (b) and $0.15$ (c).}
    \label{fig:method2}
\end{figure}

We examine the effect of the coupling strength $c$ ($= c_{12} = c_{21}$) on the inference performance. 
The proposed method performs excellently even when the coupling is not very weak (Fig ~\ref{fig:method2}a, cyan). 
This result implies that the proposed method is also effective for the non-averaged phase oscillator models.
The naive method can also accurately infer the coupling strength when the coupling is sufficiently weak ($c= 0.02$). 
The reason is that the naive method is based on the averaged model, which is derived from the stochastic Winfree model under the assumption that the coupling strength is sufficiently weak (see \nameref{appendix:averaging} for the derivation). 
When the coupling strength was adequately small, the averaged model successfully reproduced the dynamics of the true model (Fig ~\ref{fig:method2}b). 
However, as the coupling strength increased, the naive method failed to accurately infer the coupling strength and overestimated it (Fig ~\ref{fig:method2}a, magenta). For instance, when the coupling strength was set to $c=0.15$, the naive method incorrectly inferred the coupling strength to be $81$ times greater than the actual value. 
One reason for the failure of the naive model is that the averaged model cannot reproduce the dynamics of the true model (Fig ~\ref{fig:method2}c). There is a discrepancy of $O(\epsilon)$ between the non-averaged model (\eqref{eq:phase-eq-nonaveraged}) and the averaged model (\eqref{eq:noisy-kuramoto-coupling-averaged}), where $\epsilon$ is the scale of coupling strength (see Sec.~\ref{method:derive-asympt-phase-eq} for details).

\subsection{Applications}
\label{sec:applications}

In Sec.~\ref{sec:proposed}, we have demonstrated that
the proposed method can accurately infer the coupling network of oscillators described by the phase models (Eqs.~\ref{eq:phase-eq-nonaveraged} and ~\ref{eq:noisy-kuramoto-coupling-averaged}).
%
Here, we test the validity of the proposed method using the data obtained from limit-cycle oscillators, which are more realistic models compared to the phase models. 
In particular, we consider two models: the Brusselator model for chemical oscillators and the model of circadian oscillators in the suprachiasmatic nucleus (SCN).

\subsubsection{Brusselator oscillators}

\begin{figure}[h]
    \centering
    \includegraphics[width=.8\hsize]{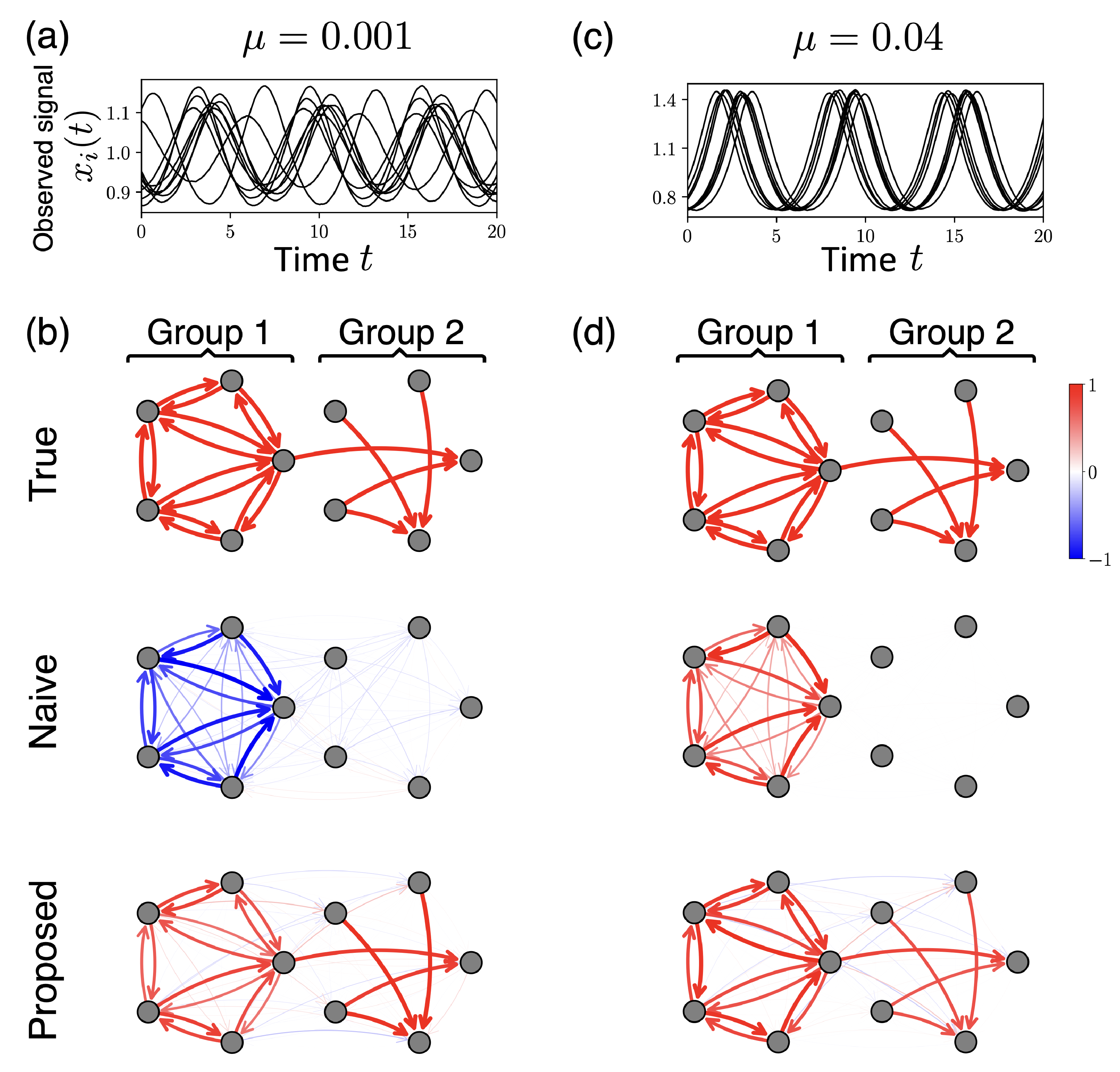}
    \caption{
    Network inference from oscillatory signals: Application to simulated data from Brusselator oscillators. \\
    (a): Observed signals $x_i(t)$ ($i=1,2, \cdots, 10$) from Brusselator oscillators whose bifurcation parameter is close to Hopf bifurcation point: $\mu=0.001$. 	
	(b): True network between Brusselator oscillators (top) and networks inferred from the observed signals (a) by the naive method (middle) and the proposed method (bottom). 
    (c): Same as (a), but the data is simulated with the bifurcation parameter that is not very close to the Hopf bifurcation point: $\mu=0.04$.  
    (d): Same as (b), but the network was inferred from the observed signals (c).     
    In (b) and (d), the relative weights of the couplings are represented by the color and width of the arrows.
    }
    \label{fig:brusselator}
\end{figure}

The first example is the Brusselator model~\cite{prigogine1971thermodynamic}, 
a two-dimensional dynamical system that describes a type of autocatalytic chemical reaction (see Sec.~\ref{method:models} for details). 
Here, we consider a network of 10 oscillators, divided into two groups: a densely connected population (Fig \ref{fig:brusselator}: group 1) and a sparsely connected population (Fig \ref{fig:brusselator}: group 2).

From the observed signal $x_i(t)$ ($i=1,2, \cdots, 10$), we infer the coupling network $\{c_{ij} \}$ ($i, j=1,2, \cdots,10$) of the averaged phase model (\eqref{eq:noisy-kuramoto-coupling-averaged}), which is expected to be approximately proportional to the coupling in the Brusselator oscillators. 
The sampling interval of the observed signal is set to $h=0.01$, and the observation duration is set to $Kh= 3.0 \times 10^5$  (approximately $5\times 10^4$ periods). 
The proposed method is compared with the naive method that infers the interaction network by fitting parameters of the averaged model (\eqref{eq:noisy-kuramoto-coupling-averaged}) (see \nameref{appendix:naive-method} for details).
In the naive method, the sampling interval is set to $h=0.1$ because otherwise the computational cost will be enormous. It was confirmed that the results are quantitatively preserved from the case of $h= 0.01$ using several shorter data sets. 

First, we consider the case where the oscillators are rather close to the Hopf bifurcation point by setting a Hopf bifurcation parameter $\mu = 0.001$. 
The waveform of the observed signal is close to a sinusoidal wave, but the amplitude of the oscillators varies due to the heterogeneity (Fig ~\ref{fig:brusselator}a). 
Fig ~\ref{fig:brusselator}b compares the true network (top) with the inferred results from the observed signals using the naive method (middle) and the proposed method (bottom). 
The naive method fails to determine the true network structure and relative strengths. It incorrectly identifies negative coupling strengths in the left populations when the true couplings are positive. In addition, it suggests that the right population is nearly isolated, with much weaker internal coupling strength compared to the left population. In the real network,  however, the coupling strengths of the two groups are identical, differing only in the density of connections. 
In contrast, the proposed method accurately infers the network. The correlation coefficient between the true coupling network and its estimate is $-0.77$ for the naive method and $0.96$ for the proposed method, indicating the superior inference performance of the proposed method. 
We further confirm the reliability of the inference result by generating 10 synthetic data sets with different input noise. The correlation coefficient (mean $\pm$ standard deviation) of the proposed method was $0.95 \pm 0.008$, while that of the naive method was $0.29 \pm 0.66$.

Second, we test whether the proposed method is applicable to the case when the waveform of the observed signal is distorted from a sinusoidal waveform (Fig ~\ref{fig:brusselator}c) and the coupling function $\gamma(\theta_j- \theta_i)$ deviates from the sinusoidal function. 
For this purpose, we examine the case where the oscillators are not very close to the Hopf bifurcation point by setting a Hopf bifurcation parameter $\mu = 0.04$. 
Similar to Fig ~\ref{fig:brusselator}b, Fig ~\ref{fig:brusselator}d compares the true network (top) with the inferred results using the naive method (middle) and the proposed method (bottom). 
While the naive method cannot infer the connections in the sparse network (group 2), the proposed method infers the network structure accurately. 
The correlation coefficient between the true coupling matrix and its estimate was $0.75$ for the naive method and $0.98$ for the proposed method. 
Similarly, the reliability of the inference result is validated by generating 10 synthetic data sets with different input noise. The correlation coefficient (mean $\pm$ standard deviation) of the proposed method was $0.95 \pm 0.02$, while that of the naive method was $-0.28 \pm 0.62$.
Furthermore, the bifurcation parameter $\mu$ was systematically increased to investigate the robustness of the proposed method. 
The proposed method demonstrated a high degree of inference accuracy for larger $\mu$, yielding correlation coefficients between the true and inferred coupling matrices of $0.92, 0.88$, and $0.90$, respectively, for $\mu=0.1, 0.2$, and $0.3$ (see \nameref{appendix:brusselator-largemu} for the inference result).
In summary, the results imply that the proposed method achieves a superior inference result to the naive method for a broad range of parameters.

We note that group 1 is highly synchronized in both cases $\mu=0.001$ and $\mu=0.04$, and the proposed method performs well even for synchronized oscillators. 
To show this, we quantify the degree of synchronization by a group-level Kuramoto order parameter 
$R_u =  \left\langle \left| \frac{1}{N_u} \sum_{j \in \mathcal{A}_u } e^{i \phi_j(t)} \right| \right\rangle$, 
where  $u \in \{1, 2\}$ indicates the 
groups in Fig ~\ref{fig:brusselator}b and d. 
Each group has $N_u = 5$ oscillators and the set of the oscillators' indices is denoted by $\mathcal{A}_u$. 
The group-level order parameter takes its maximum value $R_u=1$ when the oscillators in the group completely synchronize in phase, while it takes its minimum value $R_u=0$ when the oscillators' phases are uniformly distributed. 
We obtain $(R_1, R_2)=(0.97, 0.55)$ for $\mu=0.001$, and $(R_1, R_2)=(1.00, 0.62)$ for $\mu=0.04$ showing the high synchrony in group 1. 
This result demonstrates that the proposed method is able to extract the network structure from oscillatory signals even when the oscillators are in a highly synchronous state.

\clearpage
\subsubsection{Clock cells oscillators}

\begin{figure}
    \centering
    \includegraphics[width=\hsize]{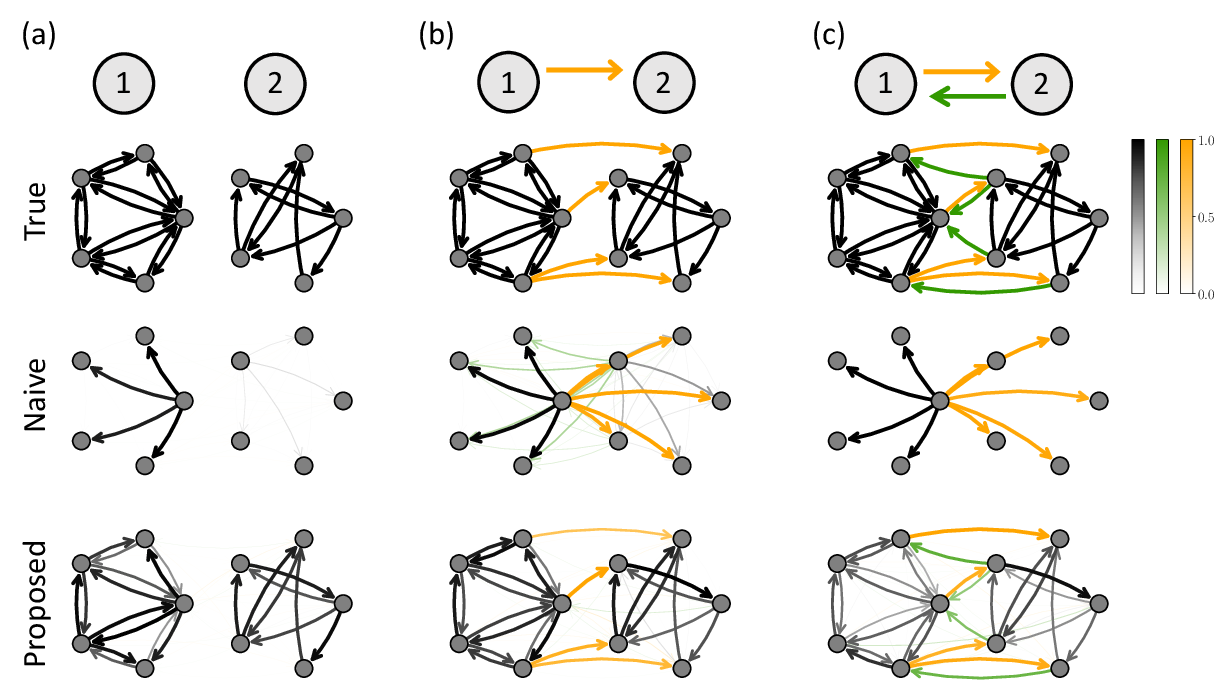}
    \caption{
    Network inference from oscillatory signals: Application to synthetic data from coupled clock cells model. \\    
	Inference results for three network structures are shown, (a): no connection between two groups, 
    (b): unidirectional connections between two groups, i.e., the interactions are from the neurons in group 1 to those in group 2, and 
    (c): bidirectional connections between two groups. 
    True interaction networks are shown in the top panels. Inferred networks from observations $y_i(t)$ (see Sec.~\ref{method:models}) by the naive method and the proposed method are shown in the middle and bottom panels, respectively. 
    The relative strength of the interactions are represented by the width of the arrows; negative weights are not shown. Intra-group connections and connections from group 1 (group 2) to group 2 (group 1) are indicated by black arrows and orange (green) arrows, respectively.     
    }
    \label{fig:clockcell}
\end{figure}

A second example is the network of clock cells in the suprachiasmatic nucleus (SCN). The mammalian circadian rhythm is driven by genetic oscillations in the clock cells, which are the neurons composing the SCN. However, it is still unclear how these clock cells interact with each other to produce a stable rhythm~\cite{MOHAWK2011349}.
In particular, the interaction between two anatomically distinct subregions of the SCN (i.e., the core and shell regions) has attracted attention because it is thought to play an important role in generating robust rhythmic signals~\cite{lu2022mammalian}.
Here, we consider a toy model of the SCN. As illustrated in Fig ~\ref{fig:clockcell}, we consider a network of model clock cells, composed of two densely connected subregions (see Sec.~\ref{method:models} for details of the model). 
Using synthetic data generated by this model, we investigate the performance of our methods in inferring the intra- and inter-networks of two subregions.

We inferred the coupling network $\{ c_{ij} \}$ ($i,j = 1,2,\cdots ,10$) from the observed signal $y_i(t)$ ($i= 1,2,\cdots ,10$) 
obtained from the clock cell model (see Sec.~\ref{method:models}). 
The sampling interval of the observation data was $h=0.04$, and the observation duration was $Kh= 1.2 \times 10^6$ (approximately $5\times 10^4$ periods). 
In the naive method, the sampling interval was $h=0.4$, 
because otherwise the computational cost will be enormous. It was confirmed that the results are quantitatively preserved from the case of $h= 0.04$ using several shorter data sets.

In this subsection, we consider three cases of interaction between groups of clock cells as follows: 
a) no interaction (Fig ~\ref{fig:clockcell}a), 
b) one-way or unidirectional interaction (Fig ~\ref{fig:clockcell}b), and 
c) two-way or bidirectional interaction (Fig ~\ref{fig:clockcell}c).  
We calculated the group-level Kuramoto order parameter $R_u$ from the clock cells in the left ($u=1$) and right ($u=2$) groups.  
The oscillators in each group are highly synchronized: the order parameter obtained from the groups 1 and 2 was $0.99$ and $0.98$ in the case of no interaction (Fig ~\ref{fig:clockcell}a), $0.99$ and $0.99$ in the case of one-way interaction (Fig ~\ref{fig:clockcell}b), and $0.99$ and $0.99$ in the case of two-way interaction (Fig ~\ref{fig:clockcell}c), respectively.  
Next, we obtain the phase $\theta_i(t)$ from the time series $y_i(t)$ $(i= 1,2, ..., 10)$ using the Hilbert transform and infer the coupling network.  
To evaluate the performance of the inference, we calculate the correlation coefficient 
between the true interaction network and the inferred network.
%
The correlation coefficient of the naive method and the proposed method was $0.23$ and $0.98$ in the case of no interaction (Fig ~\ref{fig:clockcell}a), $0.053$ and $0.97$ in the case of one-way interaction (Fig ~\ref{fig:clockcell}b), and $0.14$ and $0.95$ in the case of two-way interaction (Fig ~\ref{fig:clockcell}c), respectively.
This result suggests that the proposed method can infer the oscillator network much more accurately than the naive method. 

\begin{figure}
    \centering    
    \includegraphics[width=\hsize]{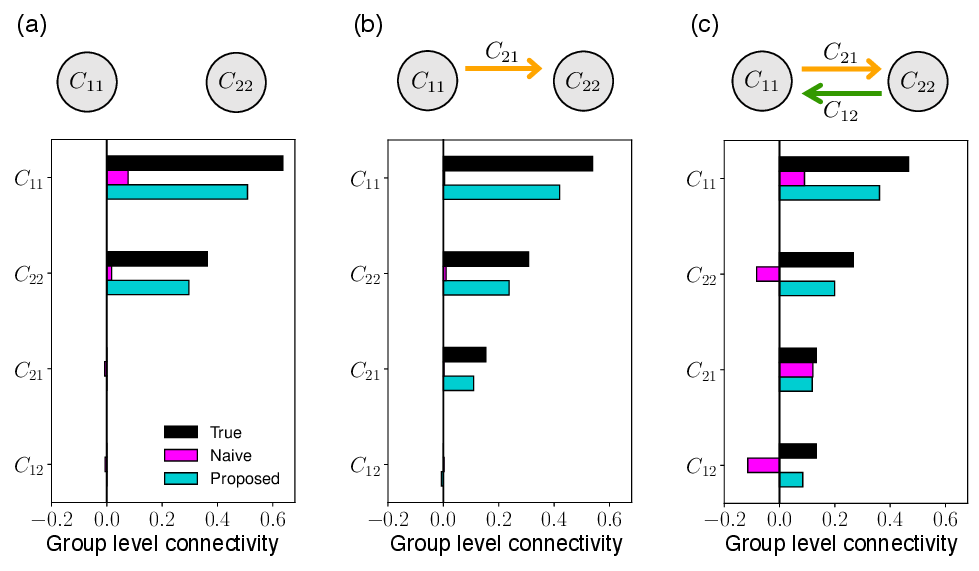}
    \caption{
    Group-level connectivity $C_{uv}$ between the clock cell groups. \\ 
    We compared the group-level connectivity from cell group $v$ to cell group $u$ of the true network with that of the inferred network by the naive and proposed methods ($u, v= 1, 2$). Same as Fig ~\ref{fig:clockcell}, we show three networks: (a) no interaction between cell groups, (b) one-way interaction between cell groups, (c) two-way interaction between cell groups. 
    }
    \label{fig:bar-ClockCell}
\end{figure}

The inference results (Fig ~\ref{fig:clockcell}) suggest that the proposed method is able to extract the inter- and intra-group interactions from the data, whereas the naive method fails to capture these interactions.   
We quantify the inter- and intra-group interactions by calculating the group level connectivity $C_{uv} = \frac{1}{|C|} \sum_{i \in \mathcal{A}_u } \sum_{j \in \mathcal{A}_v } c_{ij}$, 
where $\mathcal{A}_u$ is the set of oscillator indices in group $u \in \{ \rm 1, 2 \}$, and $|C|:= \sum_{i \in \mathcal{A}_u} \sum_{j \in \mathcal{A}_v} |c_{ij}|$ is the sum of the absolute values of inferred coupling strengths. 
Fig ~\ref{fig:bar-ClockCell} shows the group level connectivity $C_{uv}$ of the true and inferred networks. 
The inferred network by the naive method exhibits a markedly reduced level of intra-group connectivity ($C_{\rm 11}$ and $C_{\rm 22}$) compared to the true network. This result indicates that the naive method is unable to identify a densely connected group. 
In contrast, the proposed method accurately reproduces both the intra-group ($C_{\rm 11}$ and $C_{\rm 22}$) and the inter-group ($C_{12}$ and $C_{21}$) connectivity of the true network. 
These results demonstrate that the proposed method can extract not only a densely connected group of clock cells, but also the direction of interaction (e.g., one-way or two-way) between the groups.

\clearpage 

\section{Discussion}
\label{sec:concl-discuss}

We have proposed a method for inferring the network of coupled oscillators from the observed signal. 
In our approach, the phase is first reconstructed from the observed signal, and then the interaction network (i.e., coupling network) is inferred from the phase (Fig~\ref{fig:inference-protocol}). 
This method infers the coupling network $\{ c_{ij} \}$ ($i, j= 1, 2, ..., N$) of the averaged phase equation (\eqref{eq:noisy-kuramoto-coupling-averaged})  by fitting the parameters of the circle map. 
First, we have demonstrated that the proposed method can accurately infer the network when the individual oscillator dynamics is described by the phase models (Fig ~\ref{fig:method1} and ~\ref{fig:method2}).
We have subsequently validated the proposed method using simulated data from limit-cycle oscillators, i.e., the Brusselator oscillator and the clock cell model (Fig ~\ref{fig:brusselator} and ~\ref{fig:clockcell}). 
Furthermore, the result of the clock cell model (Fig ~\ref{fig:bar-ClockCell}) suggests that the proposed method can extract the group structure of the coupling network and the interaction between groups.  

Methods for inferring the coupling of oscillators can be classified into two approaches: 
(i) fitting the coupling function $q(\phi_i, \phi_j)$~\cite{kralemann2013vivo,stankovski2015coupling} and 
(ii) fitting the averaged coupling function $\gamma(\phi_i- \phi_j)$~\cite{tokuda2007inferring,panaggio2019,ota2020interaction}.
The proposed method belongs to the latter approach. The latter approach has the advantage that it can infer the coupling network and requires less data. 
However, the existing methods in this approach have limited accuracy, except when applied to a narrow class of systems, such as the Kuramoto model.  
This is because the previous methods are based on the averaged phase equation (\eqref{eq:noisy-kuramoto-coupling-averaged}), which can deviate $O(\epsilon)$ from the original dynamical system (Sec.~\ref{sec:phase-desc}). 
In contrast, the proposed method is based on a circle map that describes the phase change over one oscillation period. 
Indeed, the proposed method requires only the data with a sampling interval of the typical oscillation period $T$. However, the circle map deviates from the original system~(\eqref{eq:dyneq-noisy-coupled-osc}) by $O(\epsilon^2)$, which is more accurate than the previous methods: the deviation of $O(\epsilon)$. Consequently, our method is generally applicable to weakly coupled oscillator systems.

One advantage of the proposed method is that it is applicable to oscillators in a synchronous state. It is noted that when estimating the coupling function $q(\phi_i, \phi_j)$, the oscillators must be in a completely asynchronous state~\cite{rosenblum2001detecting, rosenblum2023inferring}. Moreover, it has been demonstrated that synchronization between oscillators deteriorates the inference accuracy of the averaged coupling function $\gamma(\phi_i- \phi_j)$~\cite{tokuda2019practical,panaggio2019}. 
Recently, it has been reported that methods based on machine learning (reservoir computing) do not provide good estimation accuracy when the oscillators are synchronized~\cite{banerjee2021machine}. 
In contrast, the proposed method can accurately estimate the coupling strength even when the oscillators are well-synchronized (Fig ~\ref{fig:method1},~\ref{fig:method2},~\ref{fig:brusselator}, and ~\ref{fig:clockcell}). The proposed method can accurately infer the network even when the Brusselator oscillators and clock cell models are strongly synchronized in the network.

The proposed method is based on three assumptions, (i) the coupling function is uniform, (ii) the coupling function is a sinusoidal function, and (iii) the observed oscillatory signal is smooth and close to the sinusoidal function.
Assumption (i) is valid for systems in which the coupling mechanism between elements is the same, including systems of cardiac cells~\cite{Alonso2016} and neurons~\cite{Vreeswijk1994,Fujita2012,Ashwin2016}. 
Note that this assumption could not be satisfied due to the diversity of neural cell types~\cite{kobayashi2009, zeng2022cell} or heterogeneity of the coupling between brain regions \cite{bastos2015dcm, safavi2023uncovering}. 
Assumption (ii) is valid when the system is in the vicinity of the Hopf bifurcation. 
As illustrated in Fig ~\ref{fig:brusselator} (c) and (d), and \nameref{appendix:brusselator-largemu}, the proposed method can accurately infer the coupling network even when the Brusselator system is not very close to the Hopf bifurcation point. This result suggests that the proposed method is effective even if the assumption (ii) is considerably violated. 
Assumption (iii) is necessary to reconstruct the phase using the Hilbert transform.  In this study, we employed the Hilbert transform, which is one of the most commonly used methods for reconstructing the phase from oscillatory signals. The results from the Brusselator and the clock cell oscillator systems demonstrate the efficacy of the proposed method in accurately inferring the network, even when the waveform deviates from a sinusoidal function.
The proposed method is effective in realistic scenarios as described above (see also Fig ~\ref{fig:brusselator} and~\ref{fig:clockcell}). However, several extensions below would dramatically increase its scope. 
As an extended model, we can consider a coupling function of the form $\gamma(\phi_i- \phi_j)= \sum_{k=1}^M a_k \cos(\phi_i- \phi_j) + b_k \sin(\phi_i- \phi_j)$. 
However, it should be noted that it would not be possible to identify the higher harmonic components when the oscillators are strongly synchronised. 
The data only provide information on the coupling function in the vicinity of the synchronised state $(\phi_i- \phi_j \approx$ const), which implies that the higher harmonic model has no practical identifiability~\cite{browning2020}. 
Additionally, it would be also an important future study to extend the proposed method to higher order interactions~\cite{casadiego2017model,bick2023higher,su2025pairwise}. 
In order to infer more complicated interactions, it is vital to incorporate a model selection approach, such as sparse regression~\cite{brunton2016discovering,nijholt2022emergent,su2025pairwise} or the likelihood ratio test~\cite{Endo2021}, into the proposed method.
Finally, it is important to note that the phase reconstruction procedure is critical in the analysis of experimental data. More advanced methods, such as the protophase-to-phase transform~\cite{kralemann2007uncovering} and the extensions of the Hilbert transform~\cite{chavez2006towards, gengel2019phase,wodeyar2021state,matsuki2023extended}, 
have the potential to further improve the performance of network inference.
In addition, it is crucial to incorporate the bandpass filtering for application involving human brain imaging data, as intra- and cross-frequency couplings are fundamental to characterizing brain activity~\cite{bastos2015dcm, onojima2021state,safavi2023uncovering, yeh2023cross}.  

In this study, we have proposed a method for inferring the coupling network from time series data measured from oscillators. 
The network inferred here is the coupling between oscillators in a reduced phase oscillator model. The inferred coupling network will provide important insights into the synchronization mechanism of oscillators. 
Moreover, unlike correlations between time series, this network captures causal relationships based on the phase information. This network is regarded as an example of effective connectivity in the field of neuroscience~\cite{Friston1994, messe2015closer, barack2022call, Kobayashi2024}.
Consequently, the application of the proposed method to oscillatory signals obtained from various systems (e.g., SCN~\cite{myung2015gaba} and spinal cord~\cite{kobayashi2016}) would offer valuable insights into the mechanisms of synchronization and information flow in the brain. 
Another promising avenue of research is to study critical phenomena in the neural system, which emerges on the edge of synchronization~\cite{botcharova2014markers, chalk2016neural, di2018landau, safavi2024signatures}, based on the networks inferred from neuronal signals.
Finally, motion control represents a promising application area for the proposed method. It is essential to understand the control mechanism of gait rhythm in order to improve rehabilitation and treatment of gait disorders (e.g. Parkinson's disease). It would be an important future study to apply the proposed method to motion capture data obtained from human subjects~\cite{arai2024interlimb}.

\section{Methods}

\subsection{Phase reduction theory for coupled oscillators}
\label{method:derive-asympt-phase-eq}
We present a phase reduction theory for the coupled oscillators (\eqref{eq:dyneq-noisy-coupled-osc}). 
Details can be found in the literature~\cite{kuramoto1984chemical,winfree2001geometry,Nakao2016}.
We first consider a limit-cycle oscillator whose dynamics are given by 
\begin{eqnarray}
    \frac{d {\bm X} }{dt}  = {\bm F}({\bm X}), 	\label{eq:single_osc_dyn}
\end{eqnarray}
where $t$ is time, ${\bm X}\in \mathbb R^d$ represents the oscillator state, ${\bm F}({\bm X}) \in \mathbb R^d$ describes the oscillator dynamics.
Let ${\bm X}_0(t)$, $T$, and $\omega= 2 \pi/T$ be the orbit, the period of the limit cycle, i.e., ${\bm X}_0(t+T)= {\bm X}_0(t)$, and the natural frequency, respectively.  
For a state $\bm X$ in the basin of attraction of the limit cycle, we can define the scalar field $\Phi_{\bm F}({\bm X}): {\mathbb R}^d \rightarrow {\mathbb R}$, referred to as the phase function, such that the phase $\phi(t) = \Phi_{\bm F}({\bm X}(t))$ satisfies
\begin{eqnarray}
    \frac{d \phi }{dt}  = \omega.
     \label{eq:phase_dynamics}
\end{eqnarray}

In this study, we consider the system of $N$ weakly coupled limit-cycle oscillators given by 
\begin{eqnarray}
    \frac{d {\bm X}_i}{dt} =
	{\bm F}({\bm X}_i)+ \epsilon \bm f_i({\bm X}_i)
		+ \epsilon \sum_{j=1}^N {\bm Q}_{ij} \left({\bm X}_i, {\bm X}_j \right) + {\bm \eta_i(t)}, 	 \label{eq:dyneq-noisy-coupled-osc-Method}
\end{eqnarray}
where $\bm F({\bm X}_i)$ represents the unperturbed dynamics of a typical oscillator, $\epsilon \bm f_i({\bm X}_i)$ represents the difference of the intrinsic dynamics of oscillator $i$ from the typical oscillator, and $\epsilon {\bm Q}_{ij}({\bm X}_i, {\bm X}_j)$ represents the interaction from oscillator $j$ to oscillator $i$. Note that $\epsilon$ is a small dimensionless parameter that characterizes the degree of heterogeneity and the coupling strength (i.e., the interaction strength). 
The noise term ${\bm \eta_i(t)}$ is assumed to be the Gaussian white noise obeying ${\rm E}[ {\bm \eta_i(t)}]= {\bm 0}$ and ${\rm Cov}[ {\bm \eta_i(t)} {\bm \eta_i(s)}]= v_i^2 I_d \delta(t-s)$, 
where $I_d$ is the identity matrix of size $d$, $v^2_i$ is the noise variance, and $\delta(t)$ is the Dirac's delta function. 
In the following, we further assume that the noise variance $v_i^2$ is sufficiently smaller than $\epsilon$ and neglect the terms of $O(v_i^2)$~\cite{yoshimura2008phase,teramae2009stochastic}. All other quantities including ${\bm F}$ and ${\bm Q}_{ij}$  are $O(1)$.   

The time derivative of the phase $\phi_i = \Phi({\boldsymbol{X}_i})$ ($i=1,2,  \cdots ,N$) can be written as 
\begin{eqnarray}
    \frac{d \phi_i}{dt} 
        &=&  \left. \grad_{\boldsymbol{X}} \Phi \right|_{\boldsymbol{X} = \boldsymbol{X}_i(t)}    \cdot \frac{d\boldsymbol{X}_i }{dt}   \nonumber  \\
        &=&  \left. \grad_{\boldsymbol{X}} \Phi \right|_{\boldsymbol{X} = \boldsymbol{X}_i(t)}    \cdot \left( {\bm F}({\bm X}_i) + \epsilon \boldsymbol{f}_i(\boldsymbol{X}_i) +
        \epsilon \sum_{j=1}^N {\bm Q}_{ij} \left({\bm X}_i, {\bm X}_j \right) + \epsilon {\bm \eta_i(t)} \right)  \quad	\label{eq:phase-eq-exct1}		\\
        &=& \omega + \left. \grad_{\boldsymbol{X}} \Phi \right|_{\boldsymbol{X} = \boldsymbol{X}_i(t)} \cdot \left( \epsilon \boldsymbol{f}_i(\boldsymbol{X}_i) + \epsilon \sum_{j=1}^N {\bm Q}_{ij} \left({\bm X}_i, {\bm X}_j \right) + \epsilon {\bm \eta_i(t)} \right). 	\label{eq:phase-eq-exct2}
\end{eqnarray}
In the derivation of \eqref{eq:phase-eq-exct2} from \eqref{eq:phase-eq-exct1}, we used an expression that follows from the definition of the phase function: 
\begin{eqnarray}
    \grad_{\boldsymbol{X}} \Phi  \cdot {\boldsymbol{F}}(\boldsymbol{X}) 
        = \omega.
\end{eqnarray}	
Suppose that the perturbations are small and the orbit $\bm X(t)$ stays in the vicinity of the limit-cycle orbit $\boldsymbol{X}_0(t)$, i.e., $\left\| \bm X(t)- \boldsymbol{X}_0(t) \right\| = O(\epsilon)$, we can derive the phase equation 
\begin{eqnarray}
    \frac{d\phi_i}{dt} = \omega 
    + {\bm Z}(\phi_i) \cdot 
         \left( \epsilon  	{\bm f}_i(\boldsymbol{X}_0 (\phi_i))  
        + \epsilon  \sum_{j=1}^N 	{\bm Q}_{ij}(\boldsymbol{X}_0 (\phi_i),  \boldsymbol{X}_0 (\phi_j) )          
        + \epsilon {\bm \eta}_i(t)
        \right) + O(\epsilon^2),	 	
     \qquad		\label{eq:phase-reduc-Z-jp}
\end{eqnarray}
where ${\bm Z}(\phi) :=  \left. \grad_{\boldsymbol{X}} \Phi \right|_{\boldsymbol{X} = \boldsymbol{X}_0(\phi / \omega)}$ is the phase sensitivity function. 
Finally, we can derive the phase equation (\eqref{eq:phase-eq-nonaveraged}) from \eqref{eq:phase-reduc-Z-jp} by introducing the following functions: $\nu_i(\phi_i):={\bm Z}(\phi_i) \cdot {\bm f}_i(\boldsymbol{X}_0 (\phi_i))$ and $q_{ij}(\phi_i, \phi_j):= {\bm Z}(\phi_i) \cdot  {\bm Q}_{ij}(\boldsymbol{X}_0 (\phi_i),  \boldsymbol{X}_0 (\phi_j) ) $. 

By averaging approximation, \eqref{eq:phase-eq-nonaveraged} can be reduced to \eqref{eq:noisy-kuramoto-coupling-averaged} with 
\begin{eqnarray}
	\omega_i &=& \omega + \frac{ \epsilon}{2\pi }   \int_0^{2\pi} \nu_i(\phi) d\phi,	\quad  \\
	c_{ij} &=&\frac{\epsilon}{\sqrt{\pi} } \left\| \frac{1}{2\pi} \int_0^{2\pi} q_{ij}( \psi, \psi+ \phi ) d\psi \right\|, \quad \\  
    \gamma_{ij}( \phi ) &=& \sqrt{\pi} \frac{  \int_0^{2\pi}  q_{ij}( \psi,  \psi+ \phi ) d\psi  }{ \left\| \int_0^{2\pi}  q_{ij}( \psi,  \psi+ \phi ) d\psi \right\| }, \quad ({\rm if} \ c_{ij}\neq 0) 
    \\  
    \sigma^2_i &=& \frac{v_i^2}{2\pi} \int_0^{2\pi} \boldsymbol{Z}^{\rm T}(\phi) \boldsymbol{Z}(\phi) d\phi,  
\end{eqnarray}
which are the natural frequency of oscillator $i$, the coupling strength from oscillator $j$ to $i$, the averaged coupling function, and the noise variance, respectively.  
 $\displaystyle \| f(\phi) \| := \left(\int_0^{2\pi} (f(\phi))^2 d\phi \right)^{1/2}$ is the $L_2$ norm of the function $f(\phi)$.
We should note the difference between the phase of the non-averaged equation (\eqref{eq:phase-eq-nonaveraged}) and the phase of the averaged equation (\eqref{eq:noisy-kuramoto-coupling-averaged}).
To avoid confusion, 
we write $\phi_i$ and $\phi_j$ in the averaged equation (\eqref{eq:noisy-kuramoto-coupling-averaged}) as $\varphi_i$ and $\varphi_j$, respectively. 
Neglecting the noise term, the averaged equation (\eqref{eq:noisy-kuramoto-coupling-averaged}) can be obtained from a variable transformation, so-called the near-identity transformation:
\begin{equation}
   \varphi_i = \phi_i + \epsilon h_i(\phi_1, \ldots, \phi_N). 
   \label{eq:near-identity}
\end{equation}
where $h_i$ is a $2\pi$-periodic function of $O(1)$ \cite{kuramoto1984chemical, hoppensteadt1997, kori2009collective, Nakao2016}. 
Equation \ref{eq:near-identity} implies that these phases are actually different by $O(\epsilon)$. 
This discrepancy could lead to a critical degradation in the inference performance of the naive method (e.g. Fig ~\ref{fig:method2}, \ref{fig:brusselator}, \ref{fig:clockcell}).

\subsection{Parameter inference by maximum likelihood method}
\label{method:mle}  
The proposed method determines the parameters of the circle map (\eqref{eq:circle-map}), 
$\alpha, \{ \omega_i \}, \{ c_{ij} \}$, and $\{ \sigma_i \}$ 
($i, j= 1, 2, \cdots, N$), using the maximum likelihood method. 
The log-likelihood function is given as 
\begin{eqnarray}
    l(\alpha, \boldsymbol{\theta}_1, \boldsymbol{\theta}_2,  \cdots , \boldsymbol{\theta}_N) = \sum_{i=1}^N l_i (\alpha, \boldsymbol{\theta}_i), \label{eq:l-sum-li}  
\end{eqnarray}
where,  
\begin{eqnarray}   
    l_i (\alpha, \boldsymbol{\theta}_i) &=& -\frac{M}{2} \log  \left(2 \pi \sigma_i^2 T \right) \nonumber\\
    && -  \frac{1}{2 \sigma_i^2 T }\sum_{m=0}^{M-1}	\left( \Delta \Phi_{i, m} - T \left\{ \omega_i + \sum_{j=1}^N c_{ij} \sin \left( \Phi_{j, m} - \Phi_{i, m} + \alpha \right) \right\} \right)^2,   \quad  \quad  \quad		\label{eq:li-def-prop}
\end{eqnarray}
and 
$M$ is the number of cycles, 
$\boldsymbol{\theta}_i := (\omega_i, c_{i1}, c_{i2}, ..., c_{iN}, \sigma_i)$ is a set of parameters associated with oscillator $i$. 
Due to the parameter $\alpha$ in \eqref{eq:li-def-prop}, the log-likelihood function is not quadratic in terms of the parameters, rendering its maximization nontrivial. 
However, for a fixed value of $\alpha$, we can maximize the function $l_i(\alpha, \boldsymbol{\theta}_i)$ by solving a set of linear equations to find the optimal parameter set: 
\begin{eqnarray}
    \hat{\boldsymbol{\theta}}_i (\alpha) = \mathop{\arg \max}_{ \boldsymbol{\theta}_i } l_i(\alpha,  \boldsymbol{\theta}_i).  \label{eq:opt_theta_i}
\end{eqnarray}
Thus, the maximization of the log-likelihood function (\eqref{eq:l-sum-li}) can be reduced to an optimization problem of the scalar function of one variable: 
\begin{eqnarray}
    l(\alpha,  \hat{\boldsymbol{\theta} }_1( \alpha), \hat{ \boldsymbol{\theta} }_2(\alpha ),  \cdots  \hat{ \boldsymbol{\theta} }_N(\alpha ) ) 
    	= \sum_{i=1}^N l_i(\alpha, \hat{\boldsymbol{\theta}}_i ( \alpha)). \label{eq:l_of_alpha}
\end{eqnarray} 
The optimization of the log-likelihood function can be efficiently solved by using the Brent method~\cite{brent2013algorithms} in a range of $-\pi/2 < \alpha \leq \pi/2$. Then, by substituting the optimal parameter $\hat{\alpha}$ into \eqref{eq:opt_theta_i}, we obtain the maximum likelihood estimator $ \{ \hat{\alpha}, \hat{\boldsymbol{\theta}}_i (\hat{\alpha}) \}$ $(i=1,2, \cdots ,N)$. 
The parameter estimation procedure is summarized as follows: 
\begin{itemize}
    \item[(i)]  Find $\alpha = \hat{\alpha}$ that maximizes the log-likelihood $l(\alpha,  \hat{\boldsymbol{\theta} }_1( \alpha), \hat{ \boldsymbol{\theta} }_2(\alpha ),  \cdots  \hat{ \boldsymbol{\theta} }_N(\alpha ) )$ using the Brent method within a range of $-\pi/2 < \alpha \leq \pi/2$.
    \item[(ii)] Obtain the maximum likelihood estimate of the parameters $ \{  \boldsymbol{\theta}_i \}$ $(i=1,2, \cdots ,N)$ by substituting $\alpha=\hat{\alpha}$ into \eqref{eq:opt_theta_i}. 
\end{itemize}

\subsection{Models of coupled limit-cycle oscillators }
\label{method:models}  
We validate the proposed method using synthetic data from coupled limit-cycle oscillators. Specifically, we used two models: Brusselator oscillators and clock cells oscillators. 

\subsubsection*{Brusselator oscillators} 

We consider a network of 10 Brusselator oscillators~\cite{prigogine1971thermodynamic} described by
\begin{eqnarray}
	\frac{dx_i}{dt} &=& A_i + x_i^2 y_i - (B_i+1) x_i + \sum_{j=1}^N K_{ij} (x_j - x_i) + \rho \xi^{(x)}_{i}(t),		\label{eq:brusselator1}\\
	\frac{dy_i}{dt} &=& B_i x_i - x_i^2 y_i + d \sum_{j=1}^N K_{ij} (y_j - y_i) + \rho \xi^{(y)}_{i}(t), 			\label{eq:brusselator2}
\end{eqnarray}
where $x_i(t)$ and $y_i(t)$ represent the state of oscillator $i$, $K_{ij}$ represents the coupling strength from oscillator $j$ to oscillator $i$, $\xi^{x, y}_{i}(t)$ are the independent Gaussian white noise with mean $0$ and variance $1$. 
In order to model the heterogeneity between oscillators, the parameter $A_i$ was drawn from the uniform distribution $[0.9999, 1.0001]$ and $B_i$ was set to $B_i= (1+\mu)(1+A_i^2)$, where $\mu$ is the Hopf bifurcation parameter. In this case, each unit behaves as a limit-cycle oscillator when $\mu > 0$. The other parameters are set as follows: $d= 1.25$ and $\rho= 0.002$. 
Here, we consider a network composed of two groups of oscillators: a densely connected population (Fig \ref{fig:brusselator}b, d: group 1) and a sparsely connected population (Fig \ref{fig:brusselator}b, d: group 2).
The coupling strength, $K_{ij}$, is set to $K_{ij}=0.001$ if there is a directed edge from oscillator $j$ to $i$, and $K_{ij}=0.0$ otherwise. 
The simulated data were generated using the Euler-Maruyama method with a time step of $0.01$. 

\subsubsection*{Clock cells oscillators}

We consider a toy model of the suprachiasmatic nucleus (SCN) composed of two densely connected subregions (Fig ~\ref{fig:clockcell}).  
We adopted the clock cell model~\cite{locke2008global} given by  
\begin{eqnarray}
    \frac{dx_i}{dt} &=& 
    \tau_i \left( v_1 \frac{K_1^n}{K_1^n+ z_i^n} - v_2 \frac{x_i}{K_2 + x_i} + v_c \frac{F_i}{K_c + F_i} \right) + \rho \xi_i^{x}(t), 	\label{eq:clock_cell_mod1}	\\
    \frac{dy_i}{dt} &=& \tau_i\left(  k_3 x_i - v_4 \frac{y_i}{K_4 + y_i} \right) + \rho \xi^{y}_i(t), 	\label{eq:clock_cell_mod2}   \\
    \frac{dz_i}{dt} &=& \tau_i\left( k_5 y_i - v_6 \frac{z_i}{K_6 + z_i}  \right) + \rho \xi^{z}_i(t), 	\\
    \frac{dr_i}{dt} &=& \tau_i\left( k_7 x_i - v_8 \frac{r_i}{K_8 + r_i}  \right) + \rho \xi^{r}_i(t),		\label{eq:clock_cell_mod4}	
\end{eqnarray}
where, $x_i$, $y_i$, $z_i$, and $r_i$ (${\rm nM}$) represent the concentration of mRNA, a clock protein, a transcriptional repressor, and a neuropeptide of clock cell $i$, respectively, and $\xi_i^{x, y, z, r}(t)$ represents the independent Gaussian white noise. 
The interactions between clock cells are mediated by the neurotransmitter, which is mathematically described as follows~\cite{kori2012structure}:
\begin{eqnarray}
	F_i = \sum_{j=1}^N A_{ij} r_j,		
\end{eqnarray} 
where $F_i$ represents an average neurotransmitter level, and
$A_{ij}$ represents the coupling strength from clock cell model $j$ to $i$. 
The coupling strength is set to $A_{ij} = 0.01$ when there is a coupling from the clock cell model $j$ to $i$; otherwise, $A_{ij} = 0.0$. 
The self-coupling was set to $A_{ii}=0.9$ ($i=1, 2, \cdots, 10$) to allow the model to oscillate autonomously even in the absence of coupling. 
Other parameters are summarized in Table~\ref{tab:ClockCellModel_parameter}. 
The simulated data were generated using the Euler-Maruyama method with a time step of $0.04$.

\begin{table}[t]
\centering
\caption{\label{tab:ClockCellModel_parameter} Parameters of the clock cell model. 
Further parameters: the Hill coefficient $n= 5$, and  the scaling factor $\tau_i$ drawn from the uniform distribution $[0.9999, 1.0001]$.} 
\vspace{0.5cm}
\begin{tabular}{lc}       \hline  
	Parameters 				&		Values            \\  	\hline  
$v_1$, $v_2$, $v_c$ \ $[{\rm nM/h}]$ \quad \quad	&	6.8355, \quad  8.4297, \quad  6.7924	\\
$v_4$, $v_6$, $v_8$ \ $[{\rm nM/h}]$ \quad \quad	&	1.0841, \quad  4.6645, \quad  3.5216	\\
$K_1$, $K_2$, $K_c$ \ $[{\rm nM}]$ \quad \quad	&	2.7266, \quad  0.2910, \quad  4.8283	\\
$K_4$, $K_6$, $K_8$ \ $[{\rm nM}]$ \quad \quad	&	8.1343, \quad  9.9849, \quad  7.4519	\\
$k_3$, $k_5$, $k_7$ \ $[{\rm /h}]$ \quad \quad		&	0.1177, \quad  0.3352, \quad  0.2282		\\
\end{tabular}
\end{table}



\section*{Acknowledgements}
We thank Arkady Pikovsky and Michael Rosenblum for helpful discussions.





\section*{Supporting information}
\subsection*{S1 Text: Derivation of the circle map}
\label{appendix:derive-cm}

We derive the circle map 
\begin{eqnarray}
	\phi_i(t+T)- \phi_i(t)= T \omega_i + T \sum_{j=1}^N c_{ij} \gamma_{ij}\left( \phi_j(t) -\phi_i(t) \right) +  \sqrt{T}\sigma_i \xi_{i, t}  + O(\epsilon^2),	 \label{eq:circle-map-cont-S1}
\end{eqnarray}
from the non-averaged phase equation 
\begin{eqnarray}
    \frac{d\phi_i}{dt} = \omega + \epsilon \nu_i(\phi_i)  + \epsilon \sum_{j=1}^N q_{ij}(\phi_i, \phi_j) + {\bm Z}(\phi_i) \cdot {\bm \eta}_i(t).
    \label{eq:phase-eq-nonaveraged-S1}
\end{eqnarray}  
Although the Stratonovich interpretation is usually employed in the literature \cite{Nakao2016,Ashwin2016}, we employ the Ito interpretation for convenience.
This treatment does not yield any significant error because the difference between these interpretations is of $O(v_i^2)$, where $v_i^2$ is the noise strength, and we are assuming $v_i^2 \ll \epsilon$ in the present paper. With the Ito interpretation, $\phi_i(t)$ and $\bm \eta_i(t)$ are not correlated.

First note that the relative phase $\psi_i(t)= \phi_i(t) - \omega t$ is a slow variable because 
$\dot \phi_i=\omega + O(\epsilon)$ in \eqref{eq:phase-eq-nonaveraged-S1}. Thus, we have
\begin{equation}
 \psi_i(s)= \psi_i(t)+ O(\epsilon) \quad \mbox{for $t<s<t+T$}.
  \label{psi}
\end{equation}
By integrating both sides of \eqref{eq:phase-eq-nonaveraged-S1} from $t$ to $t+T$
and using \eqref{psi}, we obtain
\begin{eqnarray}
    \Delta \phi_i(t) &=& \phi_i(t+T) - \phi_i(t) \nonumber \\
    &=& T \omega + \int_{t}^{t+T}  \ \left\{ \epsilon \nu_i(\phi_i(s)) + \epsilon \sum_{j=1}^N q_{ij} (\phi_i(s), \phi_j(s) ) + {\bm Z}(\phi_i(s) ) \cdot {\bm \eta}_i(s) 	\right\} ds, 	\nonumber  \\
    &=& T \omega + \int_{t}^{t+T}  \ \left\{ \epsilon \nu_i(  \omega s+ \psi_i(s) ) + \epsilon \sum_{j=1}^N q_{ij} (\omega s+ \psi_i(s), \omega s+ \psi_j(s) ) + {\bm Z}(\omega s+ \psi_i(s) ) \cdot {\bm \eta}_i(s) 	\right\} ds	\nonumber  \\
    &=& T \omega + \int_{t}^{t+T}  \ \left\{ \epsilon \nu_i(  \omega s+ \psi_i(t) ) + \epsilon \sum_{j=1}^N q_{ij} (\omega s+ \psi_i(t), \omega s+ \psi_j(t) ) + {\bm Z}(\omega s+ \psi_i(t) ) \cdot {\bm \eta}_i(s) 	\right\} ds \nonumber\\
    && + O(\epsilon^2, \epsilon v_i^2), 	\nonumber  \\
    &=& T \omega_i  + T \sum_{j=1}^N c_{ij}\gamma_{ij}\left( \phi_j(t) - \phi_i(t) \right)  + \tilde{\eta}_{i}(t) + O(\epsilon^2,\epsilon v_i^2), 
    \label{eq:circle-map-int}
\end{eqnarray}
where 
\begin{eqnarray}
	\omega_i &=& \omega + \frac{ \epsilon}{2\pi }   \int_0^{2\pi} \nu_i(\phi) d\phi,	\quad  \label{eq:def_omegai-S1}\\
	c_{ij} &=&\frac{\epsilon}{\sqrt{\pi} } \left\| \frac{1}{2\pi} \int_0^{2\pi} q_{ij}( \psi, \psi+ \phi ) d\psi \right\|, \quad \label{eq:def_cij-S1}\\ 
    \gamma_{ij}( \phi ) &=& \sqrt{\pi} \frac{  \int_0^{2\pi}  q_{ij}( \psi,  \psi+ \phi ) d\psi  }{ \left\| \int_0^{2\pi}  q_{ij}( \psi,  \psi+ \phi ) d\psi \right\| }, \quad ({\rm if} \ c_{ij}\neq 0) 
    \label{eq:def_gammaij-S1} 
\end{eqnarray}
which are the natural frequency of oscillator $i$, the coupling strength from oscillator $j$ to $i$, the averaged coupling function, and the noise variance, respectively.  
 $\displaystyle \| f(\phi) \| := \left(\int_0^{2\pi} (f(\phi))^2 d\phi \right)^{1/2}$ is the $L_2$ norm of the function $f(\phi)$.
The noise term is given by
\begin{eqnarray}
    \tilde{\eta}_{i}(t) := \int_{t}^{t + T}  \ \boldsymbol{Z}(\omega s + \psi_i(t) ) \cdot \boldsymbol{\eta}_i(s) ds,   \label{eq:tilde-eta-def}
\end{eqnarray}
which is Gaussian because $\tilde{\eta}_{i}$ is a weighted sum of independent Gaussian noise. 
The mean and the variance of $\tilde{\eta}_{i}(t)$ are given as
\begin{eqnarray}
    {\rm E}[\tilde{\eta}_{i}(t) ] = 
    {\rm E}\left[ \int_{t}^{t + T}  \ \boldsymbol{Z}(\omega s + \psi_i(t) ) \cdot \boldsymbol{\eta}_i(s) ds \right] 
    = \int_{t}^{t + T}  \  \boldsymbol{Z}(\omega s + \psi_i(t) )\cdot {\rm E}\left[ \boldsymbol{\eta}_i(s)  \right]ds = 0,  
\end{eqnarray}
\begin{eqnarray}
        {\rm Cov}[\tilde{\eta}_{i}(t), \tilde{\eta}_{j}(t+  mT) ] 
        &=& {\rm E}\left[ \int_t^{t+T} \boldsymbol{Z}(\omega s + \psi_i(t) ) \cdot \boldsymbol{\eta}_i(s) ds \int_{t+m T}^{t+(m+1)T} \boldsymbol{Z}(\omega s' + \psi_j(t+mT)) \cdot \boldsymbol{\eta}_j(s') ds'  \right] 		  \nonumber \\    
        &=& {\rm E}\left[ \int_t^{t+T} ds \int_{t+mT}^{t+ (m+1)T} ds' \boldsymbol{Z}^{\rm T}(\omega s + \psi_i(t)) \bm{\eta}_i(s)  \bm{\eta}_j^{\rm T}(s') \boldsymbol{Z}(\omega s' + \psi_j(t+ mT)) \right]     \nonumber \\  
        &=& \int_t^{t+T} ds \int_{t+mT}^{t+ (m+1)T} ds' \boldsymbol{Z}^{\rm T}(\omega s + \psi_i(t)) {\rm E}\left[ \bm{\eta}_i(s)  \bm{\eta}_j^{\rm T}(s') \right] \boldsymbol{Z}(\omega s' + \psi_j(t+ mT)) \nonumber \\
        &=& v_i^2 \delta_{ij}  \delta_{m0} \int_t^{t+T}  \boldsymbol{Z}^{\rm T}(\omega s + \psi_i(t)) \boldsymbol{Z}(\omega s + \psi_j(t) ) ds \nonumber \\
        &=& T \sigma_i^2	\delta_{ij}  \delta_{m0},
\end{eqnarray}
where $m$ is an integer
and $\sigma_i$ is defined as
\begin{eqnarray}
    \sigma^2_i &=& \frac{v_i^2}{2\pi} \int_0^{2\pi} \boldsymbol{Z}^{\rm T}(\phi) \boldsymbol{Z}(\phi) d\phi.  \label{eq:def_sigmai-S1}    
\end{eqnarray}
We denote $\tilde \eta_i(mT)$ by $\sqrt{T}\sigma_i \xi_{i,m}$, where $\xi_{i,m}$ is an independent Gaussian random variable with
$E[\xi_{i,m}]=0$ and  $E[\xi_{i,m} \xi_{j,n}]=\delta_{ij} \delta_{mn}$.
By neglecting the terms of $O(\epsilon^2, \epsilon v_i^2)$, we obtain the circle map given in \eqref{eq:circle-map-cont-S1}.

\subsection*{S2 Text: Naive method}
\label{appendix:naive-method}
\subsubsection*{Phase reconstruction by the Hilbert transform}

The Hilbert transform is a widely used method to reconstruct the phase from the observed signal $y(t)$ ~\cite{king2009hilbert-vol1}. 
First, we preprocess the signal to mitigate the Gibbs phenomenon~\cite{matsuki2023extended}: we extract the peaks from the signal and restrict the analysis from the first peak to the last peak. 
Then we reconstruct the phase using the following formula
\begin{eqnarray}
    \hat{\phi}_k := \arg\left(y_k + i H_d[ y_k ] \right),
\end{eqnarray}
where $\hat{\phi}_k $ is the reconstructed phase at time $t= kh$, $\arg(z)$ is the argument of a complex number $z$, $H_d[y_k]$ is the discrete Hilbert transform~\cite{king2009hilbert-vol1,matsuki2023extended} of the signal $\{ y_k \}$ $(k=0, 1, \cdots, K-1)$. Note that the signal $y_k= y(kh)$ is observed at $N$ time steps with a constant interval $h$.

\subsubsection*{Parameter estimation by maximizing the likelihood function}
The naive method assumes that the phase are generated from the averaged phase equation 
\begin{eqnarray}
    \frac{d \phi_i}{dt} =	\omega_i  + \sum_{j=1}^N c_{ij} \gamma_{ij}( \phi_j- \phi_i ) + \sigma_i \xi_i(t),     
    \label{eq:noisy-kuramoto-coupling-averaged-S2}
\end{eqnarray}
where $\omega_i$ is the natural frequency of oscillator $i$, $c_{ij}$ is the coupling strength from oscillator $j$ to $i$, $\gamma_{ij}(\phi)$ is the averaged coupling function, $\sigma_i^2$ is the  noise variance, and $\xi_i(t)$ is the Gaussian white noise with the mean $0$ and the variance $1$. 
As in the proposed method, we assume that the coupling is homogeneous and the coupling function is sinusoidal 
\begin{eqnarray}
    \gamma_{ij}( \phi_j - \phi_i )= \sin(\phi_j -\phi_i + \alpha),    \label{eq:gamma-S2}	
\end{eqnarray}
where $\alpha$ is a parameter that controls the phase-locked state.
Substituting \eqref{eq:gamma-S2} into \eqref{eq:noisy-kuramoto-coupling-averaged-S2} and discretizing it, we obtain
\begin{eqnarray}
        \delta \phi_{i, k} = h \left\{ \omega_i + \sum_{j=1}^N c_{ij} \sin \left( \phi_{j, k} - \phi_{i, k}+ \alpha \right) \right\} + \sqrt{h} \sigma_i \xi_{i,k},         
        \label{eq:discrete-noisy-kuramoto-S2}
\end{eqnarray}
where 
$\phi_{i, k} := \phi_i(kh) $ is the phase at time $t= kh$, 
$\delta \phi_{i, k}:= \phi_{i, k+1} - \phi_{i, k}$ is the phase change in the sampling interval $h$, 
and $\xi_{i,k}$ is the independent Gaussian random variable with mean $0$ and variance $1$. 
The naive method estimates the parameters by maximizing the log-likelihood function.

The log-likelihood function that corresponds to \eqref{eq:discrete-noisy-kuramoto-S2} can be written as follows~\cite{Lansky1983,stankovski2012inference,ota2020interaction}
:
\begin{eqnarray}
    l(\alpha, \boldsymbol{\theta}_1, \boldsymbol{\theta}_2, ..., \boldsymbol{\theta}_N) &=& \sum_{i=1}^N l_i(\alpha, \boldsymbol{\theta}_i), \\
    l_i(\alpha, \boldsymbol{\theta}_i) &=& -\frac{K}{2} \log \left(2 \pi h \sigma^2_i \right) 
    - \frac{1}{2 h \sigma_i^2} \sum_{k=0}^{K-1} \left[ \delta\hat{\phi}_{i, k} -h \left\{ \omega_i + \sum_{j=1}^N c_{ij} \sin \left( \hat{\phi}_{j, k} - \hat{\phi}_{i, k} + \alpha \right) \right\} \right]^2,    
\end{eqnarray}
where, $\boldsymbol{\theta}_i := (\omega_i, c_{i1}, c_{i2}, ..., c_{iN}, \sigma_i )$ is the vector of parameters of the $i$-th oscillator, $\hat{\phi}_{i, k}$ is the reconstructed phase of the $i$-th oscillator at time $t= kh$, and $\delta\hat{\phi}_{i, k}:= \hat{\phi}_{i, k+1}- \hat{\phi}_{i, k}$ is its time difference.

In the maximum likelihood method, the parameters $\boldsymbol{\theta}$ are obtained by maximizing the log likelihood $l(\boldsymbol{\theta})$.
Unfortunately, it is difficult to find the parameter that maximizes $l(\boldsymbol{\theta})$ in this case because the log-likelihood function $l(\boldsymbol{\theta})$ is a nonlinear function of the parameter $\alpha$. 
However, assuming $\alpha$ is known, the function $l_i(\alpha, \boldsymbol{\theta}_i)$ is a quadratic function of the parameters $\boldsymbol{\theta}_i$ and it is easy to maximize $l_i(\alpha, \boldsymbol{\theta}_i)$ 
\begin{eqnarray}
    \hat{\boldsymbol{\theta}}_i (\alpha) = \mathop{\arg \max}_{ \boldsymbol{\theta}_i } l_i(\alpha,  \boldsymbol{\theta}_i).  \label{eq:Append-hat_theta_i}
\end{eqnarray}
Note that the log likelihood is a function of a single variable $\alpha$ because it is the sum of the functions $l_i(\alpha, \hat{\boldsymbol{\theta}}_i (\alpha))$
\begin{eqnarray}
    l(\alpha,  \hat{\boldsymbol{\theta} }_1( \alpha), \hat{ \boldsymbol{\theta} }_2(\alpha ),  \cdots  \hat{ \boldsymbol{\theta} }_N(\alpha ) ) := \sum_{i=1}^N l_i(\alpha, \hat{\boldsymbol{\theta}}_i ( \alpha)) \label{eq:Append-l_of_alpha}
\end{eqnarray}
Thus, the maximum likelihood estimator $ \{ \hat{\alpha}, \hat{\boldsymbol{\theta}}_i (\hat{\alpha}) \}$ $(i=1,2, \cdots ,N)$ can be calculated by finding $\alpha$ that maximizes the log-likelihood $l(\alpha)$ using the Brent method~\cite{brent2013algorithms}  
within a range of $-\pi/2 < \alpha \leq \pi/2$
and substituting $\hat{\alpha}$ into \eqref{eq:Append-hat_theta_i}.
The parameter estimation procedure is summarized as follows: 
\begin{itemize}
    \item[(i)]  Find $\alpha = \hat{\alpha}$ that maximizes the log-likelihood $l(\alpha)$ (\eqref{eq:Append-l_of_alpha}) using the Brent method within a range of $-\pi/2 < \alpha \leq \pi/2$.
    \item[(ii)] Obtain the maximum likelihood estimate of the parameters $ \{  \hat{\boldsymbol{\theta}}_i \}$ $(i=1,2, \cdots ,N)$ by substituting $\alpha=\hat{\alpha}$ into \eqref{eq:Append-hat_theta_i}. 
\end{itemize}

\subsection*{S3 Text: Averaging approximation of the stochastic Winfree model}
\label{appendix:averaging}

We show that the stochastic Winfree model 
\begin{eqnarray}
	\frac{d\phi_1}{dt} &=& \tilde{\omega}_1 -  2 \tilde{c} \sin \phi_1 (1+ \cos\phi_2) +  \tilde{\sigma}_1 \sin \phi_1  \xi_1(t),  \label{eq:noisy-winfree1-appendix}       \\
	\frac{d\phi_2}{dt} &=& \tilde{\omega}_2 -  2\tilde{c} \sin \phi_2 (1+ \cos\phi_1) +  \tilde{\sigma}_2 \sin \phi_2  \xi_2(t).  \label{eq:noisy-winfree2-appendix} 
\end{eqnarray}  
is reduced to 
\begin{eqnarray}
        \delta \phi_{i, k} = h \left\{ \omega_i + \sum_{j=1}^N c_{ij} \sin \left( \phi_{j, k} - \phi_{i, k}+ \alpha \right) \right\} + \sqrt{h} \sigma_i \xi_{i,k},         
        \label{eq:discrete-noisy-kuramoto-S3}
\end{eqnarray}
by the averaging approximation. 
As described in Sec 4.1, a non-averaged model
\begin{eqnarray}
    \frac{d\phi_i}{dt} = \omega + \epsilon \nu_i(\phi_i)  + \epsilon \sum_{j=1}^N q_{ij}(\phi_i, \phi_j) + {\bm Z}(\phi_i) \cdot {\bm \eta}_i(t)
    \label{eq:phase-eq-nonaveraged-S3}
\end{eqnarray}  
is reduced to an averaged model 
\begin{eqnarray}
    \frac{d \phi_i}{dt} =	\omega_i  + \sum_{j=1}^N c_{ij} \gamma_{ij}( \phi_j- \phi_i ) + \sigma_i \xi_i(t),     
    \label{eq:noisy-kuramoto-coupling-averaged-S3}
\end{eqnarray}
where
\begin{eqnarray}
	\omega_i &=& \omega + \frac{ \epsilon}{2\pi }   \int_0^{2\pi} \nu_i(\phi) d\phi,	\quad  \label{eq:def_omegai}\\
	c_{ij} &=&\frac{\epsilon}{\sqrt{\pi} } \left\| \frac{1}{2\pi} \int_0^{2\pi} q_{ij}( \psi, \psi+ \phi ) d\psi \right\|, \quad \label{eq:def_cij}\\ 
    \gamma_{ij}( \phi ) &=& \sqrt{\pi} \frac{  \int_0^{2\pi}  q_{ij}( \psi,  \psi+ \phi ) d\psi  }{ \left\| \int_0^{2\pi}  q_{ij}( \psi,  \psi+ \phi ) d\psi \right\| }, \quad ({\rm if} \ c_{ij}\neq 0) 
    \label{eq:def_gammaij} \\
    \sigma^2_i &=& \frac{v_i^2}{2\pi} \int_0^{2\pi} \boldsymbol{Z}^{\rm T}(\phi) \boldsymbol{Z}(\phi) d\phi,  \label{eq:def_sigmai}    
\end{eqnarray}
and $\xi_i(t)$ is the Gaussian white noise with the mean $0$ and the variance $1$. 
The stochastic Winfree model (Eqs.~\ref{eq:noisy-winfree1-appendix} and \ref{eq:noisy-winfree2-appendix}) is a non-averaged model (\eqref{eq:phase-eq-nonaveraged-S3})
with
\begin{eqnarray}
    \omega + \epsilon \nu_i(\phi_i) &=& \tilde{\omega}_i, \label{eq:omega-winfree-appendix}\\
    \epsilon q_{ij}(\phi_i, \phi_j) &=& 
    \left\{
    \begin{array}{ll}
    -2 \tilde{c} \sin \phi_i (1 + \cos \phi_i) &  (i \neq j) \\
    0 & (i=j)
    \end{array} \right.\\
    Z(\phi_i) &=& \sin \phi_i, \label{eq:Z-winfree-appendix}\\
    \eta_i(t) &=& \tilde{\sigma}_i \xi_i(t), \\ 
    \quad v_i &=& \tilde{\sigma}_i, 
\end{eqnarray}
where the phase sensitivity function $\boldsymbol{Z}(\phi_i)$ and the noise vector $\boldsymbol{\eta}_i(t)$ are assumed to be scalars and are denoted by $Z(\phi_i)$ and $\eta_i(t)$, respectively.
We calculate the averaged frequency $\omega_i$, coupling strength $c_{ij}$, coupling function $\gamma_{ij}$, and noise $\sigma_i$.

By substituting \eqref{eq:omega-winfree-appendix} into \eqref{eq:def_omegai}, we have the averaged frequency
\begin{eqnarray}
    \omega_i = \tilde{\omega}_i.
\end{eqnarray}
For $i \neq j$, we have
\begin{eqnarray}
    \int_0^{2\pi} q_{ij}(\psi, \psi+\phi)d\psi &=& \int_0^{2\pi} - \frac{2\tilde{c}}{\epsilon} \sin \psi (1+\cos (\psi+\phi)) d\psi \nonumber \\
    &=& \frac{\tilde{c}}{\epsilon}  \int_0^{2\pi} \left\{ - 2 \sin \psi - \sin (\phi+2\psi) + \sin \phi  \right\} d\psi \nonumber \\
    &=&  \frac{2\pi \tilde{c}}{\epsilon} \sin \phi. 
\end{eqnarray}
Substituting this into Eqs.~\ref{eq:def_cij} and \ref{eq:def_gammaij}, we obtain the averaged coupling strength and the averaged coupling function, 
\begin{eqnarray}
    c_{ij} &=& \frac{\epsilon}{\sqrt{\pi}} \left\| \frac{1}{2\pi} \int_0^{2\pi} q_{ij} (\psi, \psi + \phi) \right\| 
        = \frac{\epsilon}{\sqrt{\pi}} \left \| \frac{\tilde{c}}{\epsilon} \sin \phi \right \| = \tilde{c}, \label{eq:appendix-winfree-c} \\
    \gamma_{ij}(\phi) 
        &=& \sqrt{\pi}\frac{ \int_0^{2\pi} q_{ij} (\psi, \psi + \phi) }{\left\|  \int_0^{2\pi} q_{ij} (\psi, \psi + \phi) \right\|} 
        = \sqrt{\pi} \frac{  \frac{2\pi \tilde{c}}{\epsilon} \sin \phi}{ \left\|  \frac{2\pi \tilde{c}}{\epsilon} \sin \phi \right\| } 
        = \sin \phi,  \label{eq:appendix-winfree-gamma}
\end{eqnarray}
respectively. 
We obtain the variance of the averaged noise by substituting \eqref{eq:Z-winfree-appendix}
 into \eqref{eq:def_sigmai}
\begin{eqnarray}
     \sigma_i^2 = \frac{v_i^2}{2\pi} \int_0^{2\pi} \boldsymbol{Z}^T(\phi) \boldsymbol{Z}(\phi) d\phi 
    &=& \frac{\tilde{\sigma}_i^2}{2\pi} \int_0^{2\pi} \sin^2 \phi \ d\phi = \frac{\tilde{\sigma}_i^2}{2}. \label{eq:appendix-winfree-sigma}
\end{eqnarray}
Substituting Eqs.~\ref{eq:appendix-winfree-c}--\ref{eq:appendix-winfree-sigma} into \eqref{eq:phase-eq-nonaveraged-S3}, we have the averaged equation.
\begin{eqnarray}
    \frac{d\phi_1}{dt} = \tilde{\omega}_1 + \tilde{c} \sin (\phi_2-\phi_1) + \frac{\tilde{\sigma}_1}{\sqrt{2}} \xi_1(t), \label{eq:appendix-kuramoto-1}\\
    \frac{d\phi_2}{dt} = \tilde{\omega}_2 + \tilde{c} \sin (\phi_1-\phi_2) + \frac{\tilde{\sigma}_2}{\sqrt{2}} \xi_2(t). \label{eq:appendix-kuramoto-2}
\end{eqnarray}
The discritized version of Eqs.~\ref{eq:appendix-kuramoto-1} and \ref{eq:appendix-kuramoto-2} is 
a special case of the model used in the naive method (\eqref{eq:discrete-noisy-kuramoto-S3}).

\subsection*{S1 Fig}
\label{appendix:brusselator-largemu}
{\bf Network inference from oscillatory signals: Application to synthetic data from Brusselator oscillators away from Hopf bifurcation points. }
\begin{figure}[h]
    \centering
    \includegraphics[width=\linewidth]{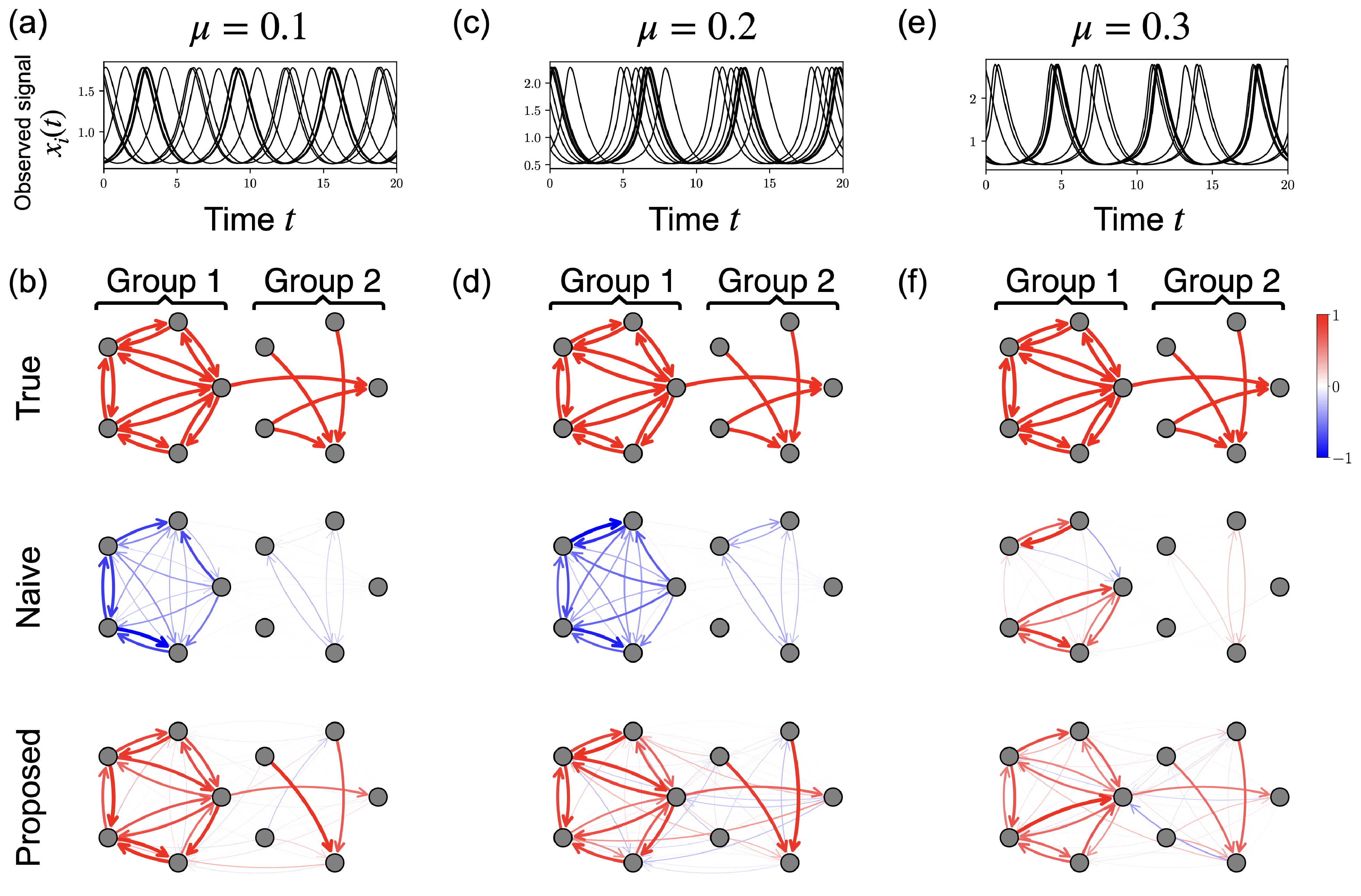}
    \caption*{
    Network inference from oscillatory signals: 
    Application to synthetic data from Brusselator oscillators away from Hopf bifurcation points. 
    The coupling networks were inferred from the synthetic data of 10 Brusselator oscillators, whose bifurcation parameter is away from the Hopf bifurcation point: $\mu=0.1, 0.2, $ and $0.3$. 
    The proposed method (bottom) is compared with the naive method (middle) that infers the coupling network by fitting parameters of the averaged model (Eq. \ref{eq:noisy-kuramoto-coupling-averaged}) (see S2 Text for details).  
    }
    \label{fig:brusselator-largemu}
\end{figure}



\end{document}